\documentclass[a4paper,fleqn,usenatbib]{mnras}

\usepackage{amsmath}	
\usepackage{amssymb}	
\usepackage{txfonts}

\usepackage[T1]{fontenc}
\usepackage{ae,aecompl}

\usepackage{graphicx}	
\usepackage{threeparttable}%
\usepackage{subfig}
\usepackage[]{hyperref}  
\hypersetup{colorlinks=true,linkcolor=blue,citecolor=blue,filecolor=blue,urlcolor=blue}

\title[A Newly-Discovered Radio Halo in MACS J2243.3-0935]{A Newly-Discovered Radio Halo in Merging Cluster MACS J2243.3-0935}

\author[T. M. Cantwell et al.]{
T. M. Cantwell$^{1}$\thanks{E-mail: therese.cantwell@postgrad.manchester.ac.uk},
A. M. M. Scaife$^{1}$, N. Oozeer$^{2,3,4}$, Z. L. Wen$^{5}$, J. L. Han$^{5}$
\\
$^{1}$Jodrell Bank Centre for Astrophysics, Alan Turing Building, Oxford Road, Manchester M13 9PL, UK.\\
$^2$SKA South Africa, The Park, Park Road, Pinelands, Cape Town 7405, South Africa.\\
$^3$African Institute for Mathematical Sciences, 6-8 Melrose Road, Muizenberg 7945, South Africa\\
$^4$Centre for Space Research, North-West University, Potchefstroom 2520, South Africa.\\
$^{5}$National Astronomical Observatories, Chinese Academy of Sciences, 20A Datun Road, Chaoyang District, Beijing 100012, China.
}

\date{Accepted XXX. Received YYY; in original form ZZZ}

\pubyear{2015}

\begin{document}
\label{firstpage}
\pagerange{\pageref{firstpage}--\pageref{lastpage}}
\maketitle

\begin{abstract}
We report the discovery of a radio halo in the massive merging cluster MACSJ2243.3-0935, as well as a new radio relic candidate, using the Giant Meterwave  Radio Telescope and the KAT-7 telescope. The radio halo is coincident with the cluster X-ray emission and has a largest linear scale of approximately 0.9 Mpc. We measure a flux density of $10.0\pm 2.0\, \rm  mJy$ at 610 MHz for the radio halo. We discuss equipartition estimates of the cluster magnetic field and constrain the value to be of the order of $1\, \rm \mu G$. The relic candidate is detected at the cluster virial radius where a filament meets the cluster. The relic candidate has a flux density of $5.2\pm 0.8\, \rm  mJy$ at 610 MHz. We discuss possible origins of the relic candidate emission and conclude that the candidate is consistent with an infall relic.

\end{abstract}




\section{Introduction}

Galaxy Clusters are the largest viralised structures in the Universe with typical masses of order 10$^{15}$ M$_{\odot}$. Most of this mass is composed of dark matter. The other 10-20$\%$ is contained in baryonic matter, with the mass in the hot intra-cluster medium (ICM) being about 10 times larger than the mass contained in galaxies \citep{Kravtsov2012,BRUNETTI2014}. The ICM was first detected in the X-ray band, emitting via thermal Bremsstrahlung, indicating that the ICM is a thermalised plasma \citep{Voit2005}. However the detection of Mpc scale diffuse emission in the radio band provides evidence that cosmic ray electrons (CRe) are also present in the ICM, as are cluster-scale magnetic fields \citep{BRUNETTI2014,Feretti2012}. As such, radio observations of clusters provide a unique opportunity to study the non-thermal populations of the ICM.

The dynamics and evolution of galaxy clusters can also be indirectly probed using  radio observations. Diffuse radio emission in clusters is divided into three morphological classes: radio relics; giant radio halos; mini halos. Radio relics are normally elongated structures found at the periphery of clusters and can be highly polarised. Giant radio halos are usually found at the center of clusters and typically have a more rounded morphology than radio relics. Giant radio halos tend to be largely unpolarised due to a high degree of either beam or internal depolarisation. Both giant radio halos and radio relics have typical physical sizes of 1 Mpc. Both radio halos and radio relics are thought to be linked to cluster mergers where shocks and turbulence are expected to accelerate electrons to relativistic energies. See \citet{Feretti2012} and references therein for a review on diffuse radio emission in clusters.

Since their discovery, a number of empirical scaling relations have been found between the radio power of giant radio halos and the properties of the host cluster such as cluster mass, temperature and X-ray luminosity \citep{Colafrancesco1999,Govoni2001,Feretti2002,Ensslin2002,Feretti2003,Brunetti2009,Cassano2013,Yuan2015}. The most well studied scaling relationship is between the radio power at 1.4 GHz, $P_{1.4}$, and the X-ray luminosity, $L_{\rm x}$, of the ICM. When both clusters with a radio halo and radio quiet clusters are examined, a bimodality is found in the distribution on the radio-X-ray plane, with radio loud clusters exhibiting a correlation and radio quiet clusters showing none \citep{Cassano2013}. This bimodality is also found to correspond to the dynamical state of the cluster, further linking radio halos to cluster mergers.

More recently a correlation between the radio power of halos at 1.4 GHz and the integrated Sunyaev-Zel'dovich (SZ) effect measurements, $Y_{500}$ was reported by \citet{Basu2012}. No bimodality in the cluster distribution was seen in the sample reported by \citet{Basu2012}; however a more statistically complete study by \citet{Cassano2013} reported a bimodal distribution of clusters in the $P_{1.4}-Y_{500}$ diagrams.

MACS J2243.3-0935 is a massive galaxy cluster at a redshift of $ z=0.447$ at the center of the super cluster SCL2243-0935 \citep{Schirmer2011}. Table~\ref{table:prop} lists some important properties of MACS J2243.3-0935 determined from previous studies across a range of wavelengths \citep{Ebeling2010,PSZcatalogue,Mantz2010,Mann2012,Wen2013}. 

The dynamical state of this cluster has been examined using a variety of techniques. \citet{Mann2012} use the X-ray morphology and the offset between the brightest cluster galaxy (BCG) and X-ray peak/centroid to characterise the dynamical state of clusters. They find that the merger axis suggested by the highly elongated X-ray emission in MACS J2243.3-0935 is mis-aligned with the merger axis suggested by the two main galaxy concentrations. \citet{Mann2012} also report a separation between the BCG and the X-ray peak of 125$\pm$6 kpc and a centroid shift, $w$, of 156$\pm$4 kpc.

\citet{Wen2013} calculate the relaxation parameter, $\Gamma$, of MACS J2243.3-0935 to be $-1.53 \pm 0.07$. The relaxation parameter measures the amount of substructure in a galaxy cluster based on the cluster's optical properties. Positive values of $\Gamma$ indicate a relaxed system and negative values of $\Gamma$ indicate a disturbed cluster. The relaxation parameter separates relaxed and unrelaxed clusters with a success rate of 94$\%$. \citet{Wen2013} find a correlation between the relaxation parameter and the offset of the halo radio power from that expected from the $P_{\rm 1.4GHz}-L_{\rm x}$ relation. This provides further evidence that the dynamical state of the cluster plays an important role in the generation of a radio halo and that the relaxation parameter could prove a powerful tool for identifying possible host clusters of diffuse radio emission. The highly negative relaxation parameter of MACS J2243.3-0935, as well as its high luminosity, were the primary reasons for selecting this cluster for study in this work. 

MACS J2243.3-0935 has also been detected by \textit{Planck} as PSZ2 G056.93-55.08 \citep{Planck2015}. From these data, the total mass measured from the SZ effect for this cluster is 1.007$\times 10^{15}$ M$_{\odot}$ and the cluster has an integrated Compton-$y$ parameter, $Y=16.3\pm1.7$ arcsec$^{2}$. In the radio, the cluster field was observed by the NRAO VLA Sky Survey (NVSS) \citep{Condon1998A} and the Faint Images of the Radio Sky at Twenty-cm survey (FIRST) \citep{Becker1995} however the field is not covered by the Sydney University Molonglo Sky Survey (SUMSS) \citep{Bock1999} or the Westerbork Northern Sky Survey (WENSS) \citep{Rengelink1997}. 

In this paper we present new observations of MACS J2243.3-0935 at 1.4 GHz using the Karoo Array Telescope\footnote{For more information see \url{http://public.ska.ac.za/kat-7}} (KAT-7) and at 610 MHz with the Giant Meter Wave Telescope (GMRT) \citep{Swarup1991}. In \S~\ref{sec:observations} we describe the observations and data reduction. In \S~\ref{sec:results} we present our results which are then discussed in \S~\ref{sec:disscusion} before making our concluding remarks in \S~\ref{sec:conclusion}.

In this work, a $\Lambda$CDM cosmology is assumed with $H_{0}=70\rm \, km\, s^{-1}\,Mpc^{-1}$, $\Omega_{m}=0.3$, $\Omega_{\Lambda}=0.7$. Using these parameters, at a redshift of 0.447, 1 arcsec corresponds to a physical scale of 5.74 kpc \citep{Wright2006}.

\begin{table}
 \centering
 \begin{threeparttable}[b]
  \caption{Properties of MACS J2243.3-0935.}
  \label{table:prop}
  \begin{tabular}{@{}lll@{}}
  \hline
 property & Value & Reference\\
 \hline
 $z$ & 0.447 & \cite{Ebeling2010}\\
 $M_{\rm 500}$ ($\times10^{14} M_{\odot}$) & 10.07$\pm$0.58 & \cite{PSZcatalogue}\\
 $Y_{\rm 500}$ (Mpc$^{2}$) & 16.3$\pm$1.7 & \cite{PSZcatalogue}\\
 $L_{\rm x,500}$ (10$^{44}$erg s$^{-1}$) & 11.56$\pm$0.67 & \cite{Mantz2010}\\
 $T$ (keV) & $8.24\pm 0.92$ & \cite{Mantz2010}\\
 $w$ (kpc) & 156$\pm $4 & \cite{Mann2012}\\
 $\Gamma$ & -1.53$\pm$0.07 & \cite{Wen2013}\\
Virial Radius (Mpc) & $2.13^{+0.18}_{-0.12}$ & \cite{Schirmer2011}\\
\hline
\end{tabular}
\begin{tablenotes}
 \item [] $L_{\rm x}$ is core excised from 0.1-2.4kev
\end{tablenotes}
\end{threeparttable}
\end{table} 

\section{Observations and Data Reduction}
\label{sec:observations}

MACS J2243.3-0935 was observed at 1.8GHz by KAT-7 and at 610 MHz with the GMRT.

The GMRT is an array of 30 antennas with diameters of 45m in India. 14 antenna make up the central core of the array while the remaining 16 antennas are arranged around the core to form a Y-shape. This distribution of antennas affords good $uv$ coverage on both the short and long baselines. The GMRT is capable of observing at a range of frequencies from 150MHz to 1280 MHz with a maximum bandwidth of 33MHz. At 610 MHz the maximum resolution is about 5 arcsec and the half power point of the primary beam is 43 arcmin.

KAT-7 is an array of 7 antennas with diameters of 12m in South Africa. KAT-7 can observe at central frequencies of 1382 MHz and 1826 MHz with a bandwidth of 256 MHz. The maximum resolution of KAT-7 is 4 arcmin at 1382 MHz and 3 arcmin at 1826 MHz.

A summary of the observational details can be found in Table \ref{table:ObsDets}. 

\begin{table}
 \centering
  \caption{Observation details of MACS J2243.3-0935.}
  \label{table:ObsDets}
  \begin{tabular}{@{}lll@{}}
  \hline
Telescope & GMRT & KAT-7\\
\hline
Date & 20 Jun 2014 & 7 Sep 2012 \\
Frequency (MHz) & 610 & 1826 \\
Time on Target (hrs) & 5.6 & 7.5 \\
Usable Time (hrs) & 5.6  & 7.5\\
Bandwidth (MHz) & 33  & 400  \\
Usable Bandwidth (MHz) & 29  &  256 \\
No. Channels & 256 & 600\\
No. Averaged Channels & 28 & 9\\
$\%$ flagged & 33$\%$  & 18.5$\%$\\
Sensitivity & $40\, \rm \mu Jy$ & $500\, \rm \mu Jy$ \\
Angular Resolution & $\sim 5\, \rm arcsec$ & $\sim 3\, \rm arcmin$\\
FOV & 43 arcmin  & $\sim 60$ arcmin\\

\hline
\end{tabular}
\end{table}


\subsection{KAT-7}
\begin{figure}
  \includegraphics[width=0.5\textwidth]{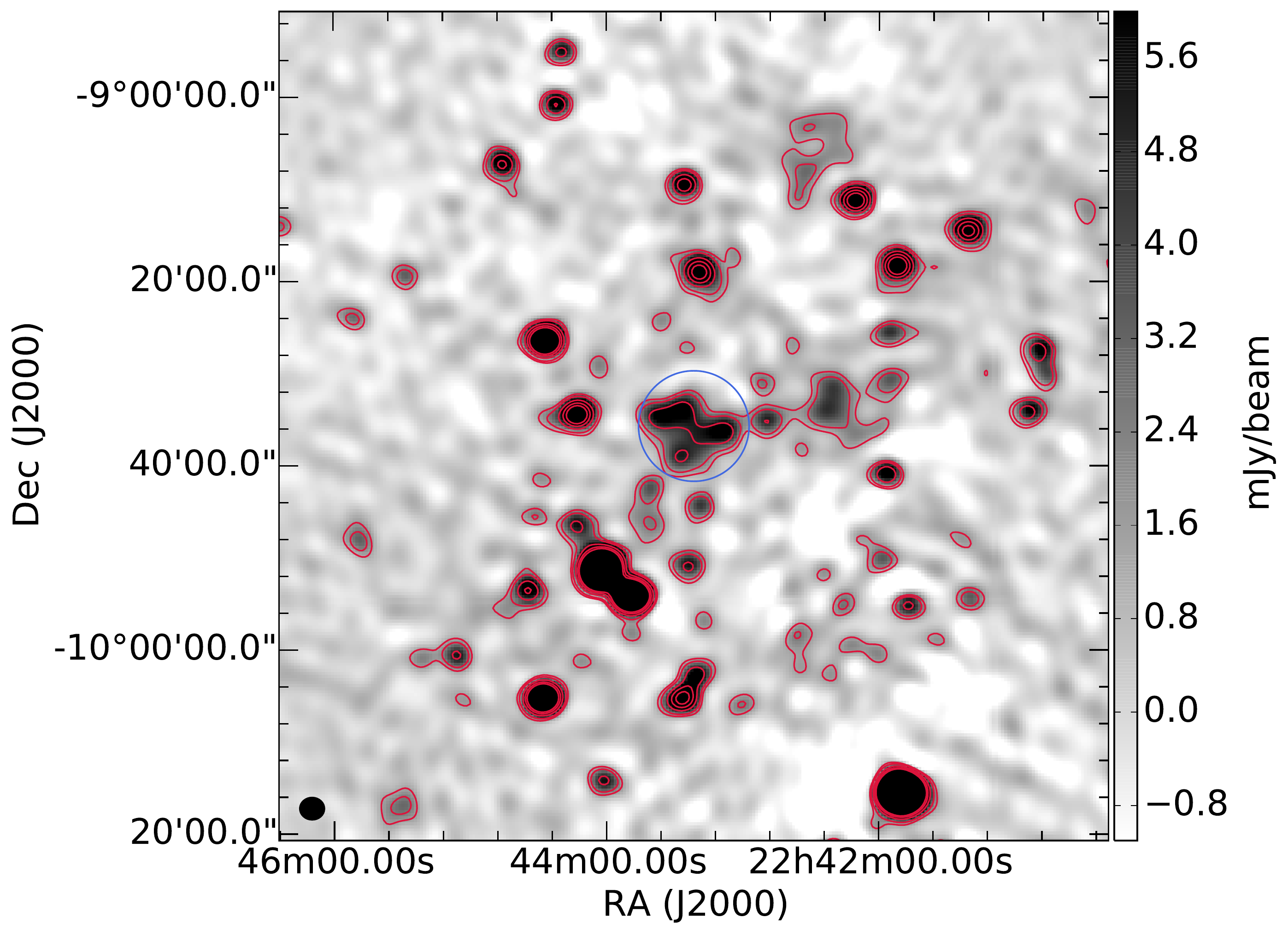}
  \caption{Greyscale plot showing the KAT-7 FOV image of MACS J2243.3-0935 with KAT-7 contours overlaid in red. Contours are at 3, 5, 10, 15, 20 $\times$ $\sigma_{\rm rms}$ where $\sigma_{\rm rms} = 500\, \rm \mu Jy/beam$. The resolution is 160.10$\times$144.99 arcsec. The blue circle marks the viral radius of MACS J2243.3-0935. The virial radius is $2.13^{+0.18}_{-0.12}\, \rm Mpc$ or 370 arcsec.  }
  \label{figure:KAT7_FOV}
\end{figure}

\begin{figure}
 \includegraphics[width=0.5\textwidth]{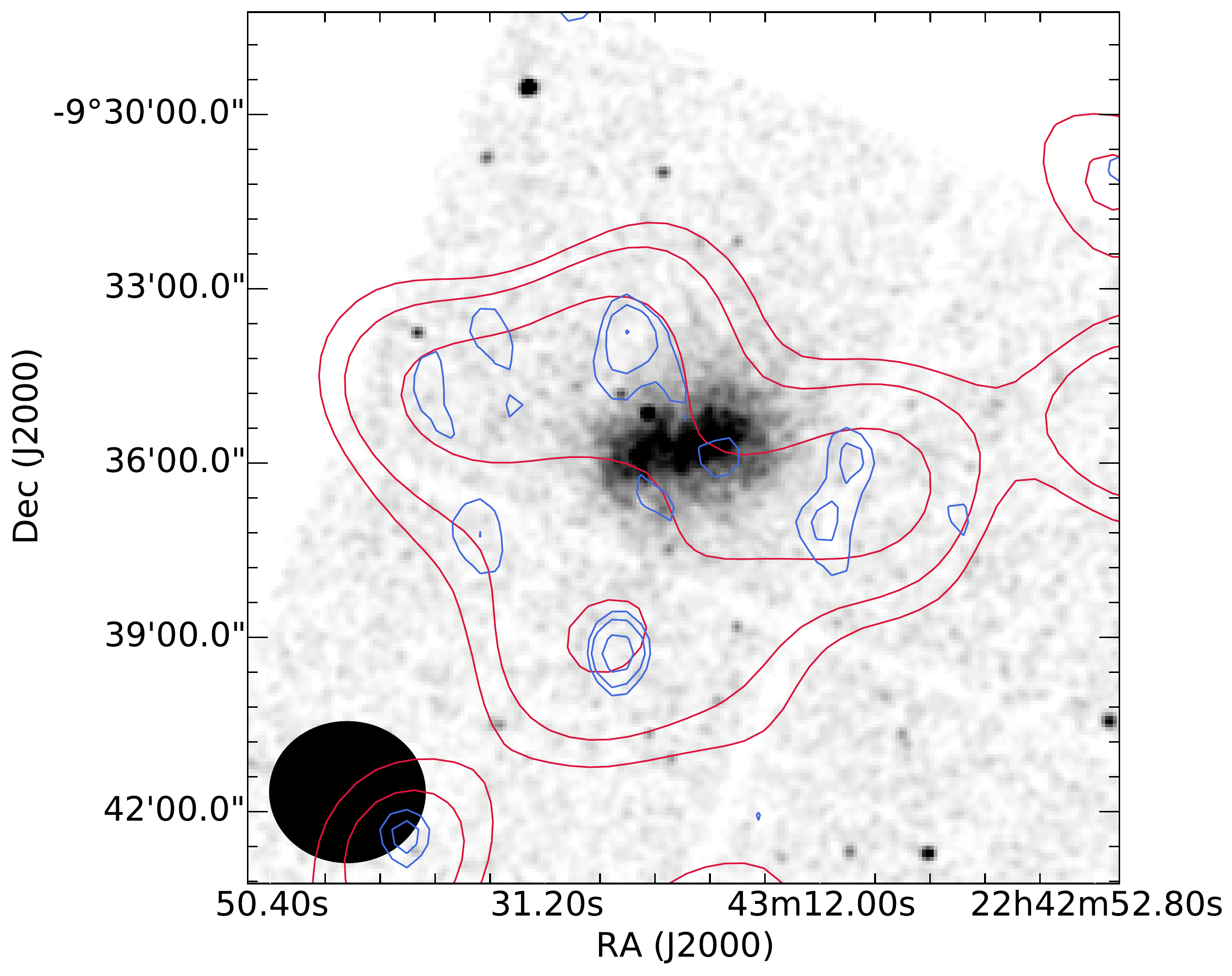}
 \caption{Greyscale plot showing X-ray image of MACS J2243.3-0935 in the Chandra ACIS $0.5-7$ keV band. The image has been smoothed with a gaussian kernel with $\sigma=\rm 3\, pixels$. KAT-7 contours are overlaid in red while NVSS contours are overlaid in blue. Contours are at 3, 5, 10, 15, 20 $\times$ $\sigma_{\rm rms}$ where $\sigma_{\rm rms} = 500\, \rm \mu Jy/beam$ for KAT-7 and $\sigma_{\rm rms} = 400\, \rm \mu Jy/beam$ for NVSS. The resolution of the KAT-7 image is 160.10$\times$144.99 arcsec while the resolution of the NVSS image is $45 \times 45\, \rm arcsec$}
 \label{figure:KAT7}
\end{figure}

KAT-7 was used to observe MACS J2243.3-0935 on the 2012-09-08. All seven antennas were used for this observation at a frequency of 1822 MHz with a bandwidth of 400 MHz. Due to analog filters in the IF and baseband system only the central 256 MHz of the bandwidth is useable. A first round of flagging was carried by the automatic flagging routine (developed in-house) to remove known radio frequency interference (RFI). The data were further flagged inside CASA to remove other low level RFI. PKS 1934-638 was used as the primary calibrator and PKS2243-123 as the phase calibrator. The flux
calibrator was observed every 2 hours for 2min while the phase calibrator was observed every 15min for 3 min. Flux densities were tied to the Perley-Butler-2010 flux density scale \citep{Perley2013}. Standard data flagging and calibration was carried out in CASA4.3. Three rounds of phase only selfcal were performed. The data were then imaged using the multifrequency, multiscale clean task in CASA with a briggs weighting robust parameter of 0. The resulting image has an rms of $0.5\, \rm mJy$. Figure~\ref{figure:KAT7_FOV} shows the full field KAT-7 image of MACS J2243.3-0935 while Figure~\ref{figure:KAT7} shows the central cluster region.

\subsection{GMRT}
MACS J2243.3-0935 was observed by the GMRT at 610 MHz with a bandwidth of 33 MHz on 20th of June 2014.\footnote{MACS J2243.3-0935 was also observed by the GMRT for 7 hours on 25th October 2010 at 610 Mz and 235 MHz. These observations were taken before upgrades began on the GMRT and are of lower quality than the new observations and are significantly contaminated with RFI. Including these data does not improve the image quality.} The primary calibrator, 3C48, was observed for fifteen minutes at the start of the observations and 3C468.1 was observed at the end of the observation in order to test the flux calibration. The phase calibrator J2225-049 was observed for 5  minutes, every 20  minutes. Data reduction and calibration for GMRT data at 610 MHz were carried out in CASA \citep{McMullin2007}. The calibration process followed that described in \citet{DeGasperin2014}. Calibration and flagging were performed in an iterative fashion. Data were phase, amplitude and bandpass calibrated then flagged, in the first round, using first the rflag mode in the CASA task flagdata and, in the second round, with AOFlagger \citep{offringa2012}. After flagging with AOFlagger, the data were averaged and a final round of calibration and flagging with AOFlagger was performed. Five rounds of phase only selfcal were carried out. After calibration and selfcal, approximately 33$\%$ of the data were flagged. Flux densities were tied to the Perley-Butler-2010 flux density scale \citep{Perley2013}. The flux density measured from 3C468.1 was $12.5\pm0.9\, \rm  Jy$, which agrees within error with the literature value of 12.7 Jy \citep{Helmboldt2008,Pauliny1966}. The data were then imaged using the multifrequency, multiscale clean task in CASA.   

\begin{figure}
   \includegraphics[width=0.5\textwidth]{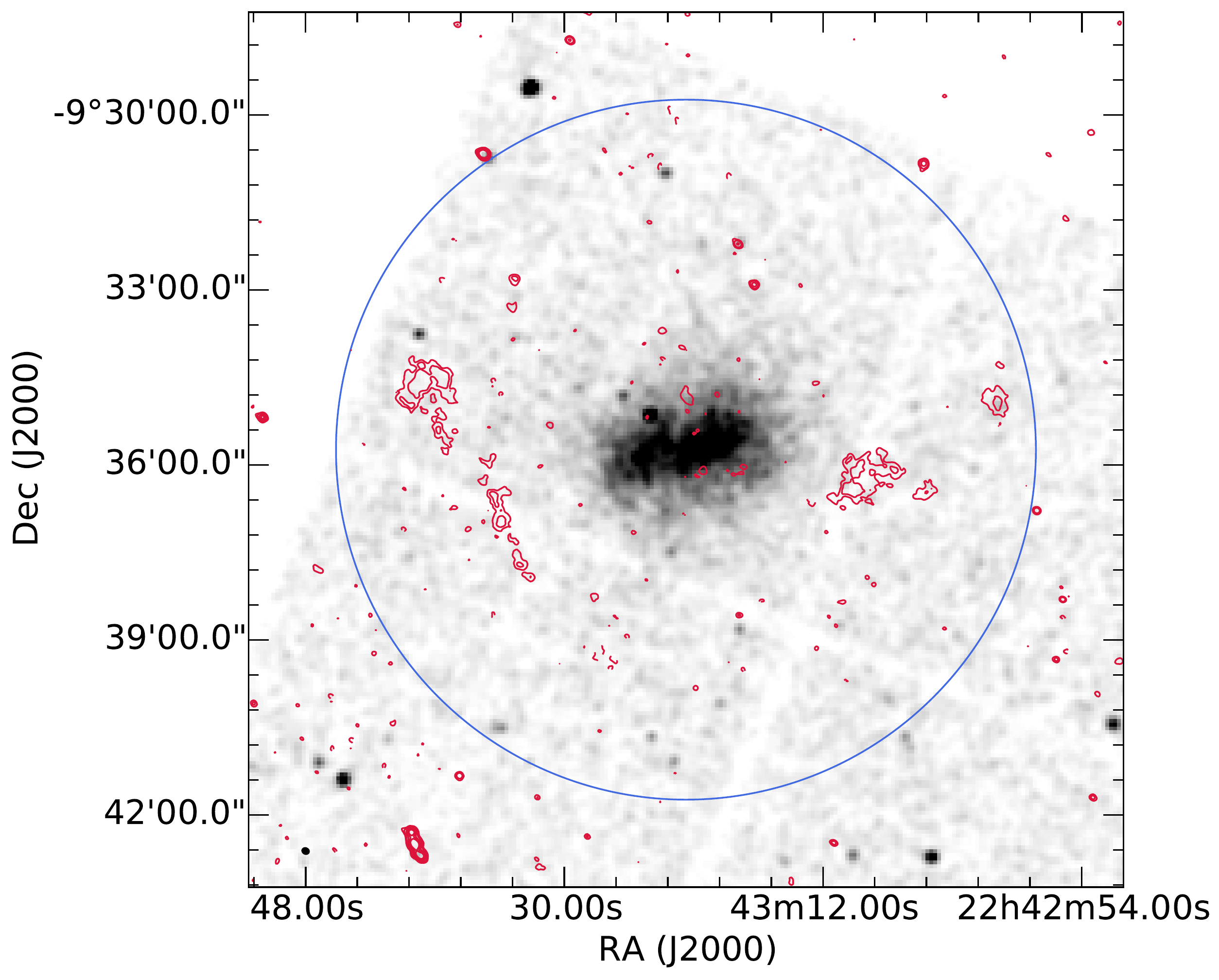}
  \caption{Greyscale plot showing X-ray image of MACS J2243.3-0935 in the Chandra ACIS $0.5-7$ keV band. The image has been smoothed with a gaussian kernel with $\sigma=\rm 3\, pixels$. The red contours show the naturally weighted GMRT image. Contours are at -3, 3, 5, 10, 15, 20 $\times$ $\sigma_{\rm rms}$ where $\sigma_{\rm rms}=45\, \rm \mu Jy/beam$. The resolution of the GMRT image is 7.44$\times$6.06 arcsec. Radio point sources have been subtracted from this image. The blue circle marks the viral radius of MACS J2243.3-0935. The virial radius is $2.13^{+0.18}_{-0.12}\, \rm Mpc$ or 370 arcsec.}
  \label{figure:highres}
\end{figure}

\begin{figure}
   \includegraphics[width=0.5\textwidth]{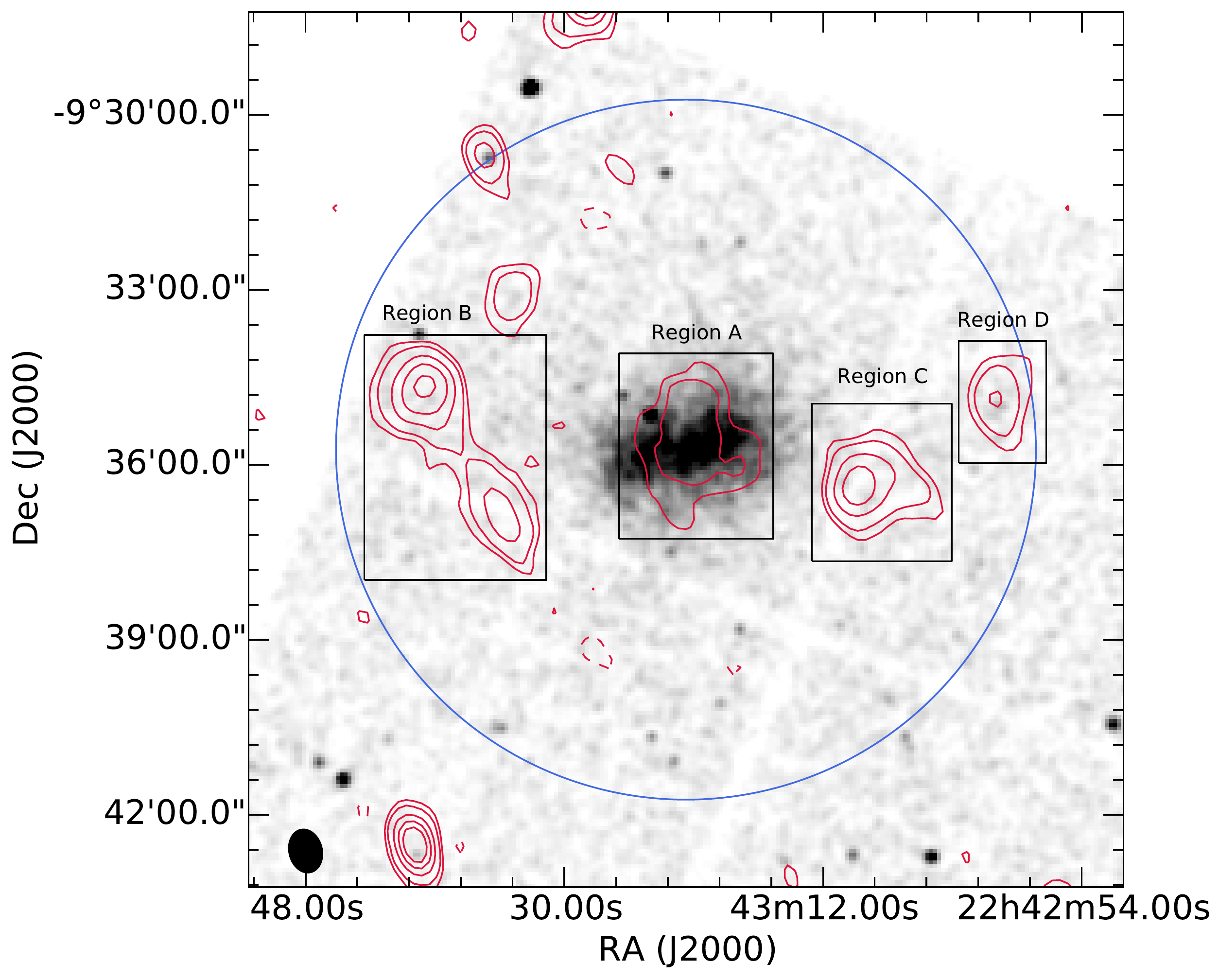}
  \caption{Greyscale plot showing X-ray image of MACS J2243.3-0935 in the Chandra ACIS $0.5-7$ keV band. The image has been smoothed with a gaussian kernel with $\sigma=\rm 3\, pixels$. The red contours show the naturally weighted GMRT image with a uvtaper of 5$\times$4 k$\lambda$. Contours are at -3, 3, 5, 10, 15, 20 $\times$ $\sigma_{\rm rms}$ where $\sigma_{\rm rms}=200\, \rm \mu Jy/beam$. The resolution of the GMRT image is $44.89\times 33.70$ arcsec. Radio point sources have been subtracted from this image. The blue circle marks the viral radius of MACS J2243.3-0935. The virial radius is $2.13^{+0.18}_{-0.12}\, \rm Mpc$ or 370 arcsec. Different regions of diffuse radio emission are marked with black rectangles and labelled.}
  \label{figure:lowres}
\end{figure}

To subtract the point source population, data from baselines longer than 4 k$\lambda$, which corresponds to an angular scale of 50 arcsec, were imaged with a Briggs robust parameter of 0. The resulting image had an rms of 40 $\mu$Jy/beam and a resolution of 4.84 $\times$ 4.15 arcsec. The clean components for compact sources in the cluster were then Fourier transformed and subtracted from the $uv$ data using the CASA tasks ft and uvsub. Two images were then made using the point source subtracted data: the first was naturally weighted with no uvtaper and the second was naturally weighted with a uvtaper of 5 k$\lambda$ by 4 k$\lambda$. The naturally weighted image has a resolution of 7.44 $\times$ 6.06 arcsec and an rms of 45 $\mu$Jy/beam and is shown in Figure~\ref{figure:highres}. The tapered image with a resolution of 44.89 $\times$ 33.70 arcsec and an rms of 200 $\mu$Jy/beam and is shown in Figure~\ref{figure:lowres}.

\section{Results}
\label{sec:results}
\subsection{MACS J2243.3-0935 at $>$ 1 GHz}
\label{sec:KAT7_results}

Figure~\ref{figure:KAT7_FOV} shows the KAT-7 image of MACS J2243.3-0935 at 1822 MHz. The resolution of the image is 160.10$\times$144.99 arcsec. However, the fidelity of the KAT-7 image is limited by the presence of a bright complex source to the south west of the field of view (FOV).  A large region of diffuse emission can be seen towards the cluster center with a largest linear scale  (LLS) of 3.9 Mpc. This diffuse emission is detected at a 5$\sigma$ level. The flux density of the cluster emission detected by KAT-7, measured within the $3\sigma$ level, is $40\pm6\, \rm mJy$. Comparison with NVSS in Figure~\ref{figure:KAT7} shows that there are at least two unresolved point sources in this region. Given the low resolution of these data it is not possible to characterise the emission in the KAT-7 image further than to note that diffuse emission appears to be present in addition to these compact sources. 

\subsection{MACS J2243.3-0935 at $<$ 1 GHZ} 
\begin{figure}
 \includegraphics[width=0.5\textwidth]{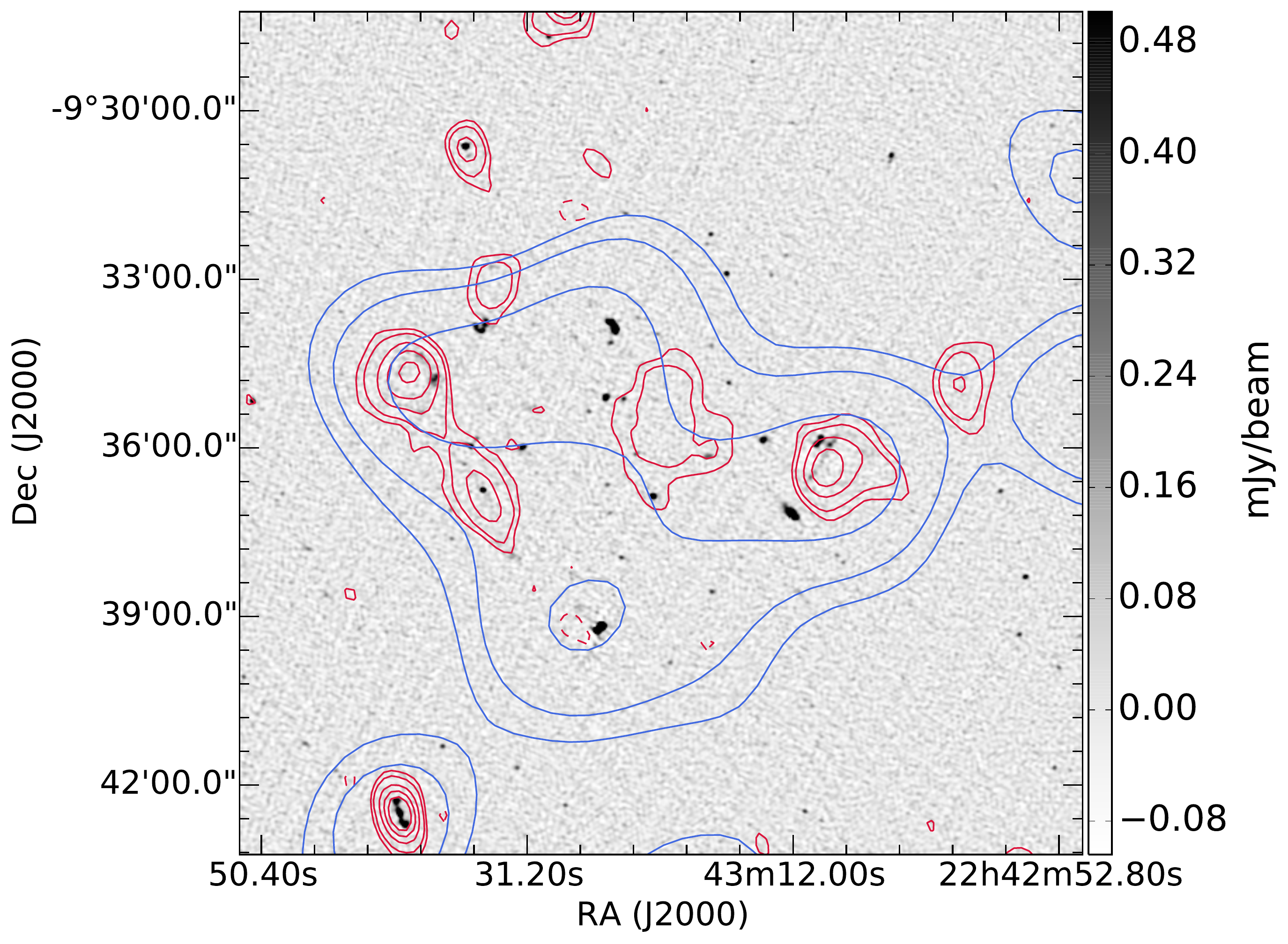}
 \caption{Greyscale image shows the robust 0 high resolution GMRT data used for the point source subtraction. KAT-7 contours are overlaid in blue while contours for the tapered, point source subtracted GMRT image are overlaid in red. Contours are at 3, 5, 10, 15, 20 $\times$ $\sigma_{\rm rms}$ where $\sigma_{\rm rms} = 500\, \rm \mu Jy/beam$ for KAT-7 and -3 3, 5, 10, 15, 20 $\times$ $\sigma_{\rm rms}$ $\sigma_{\rm rms} = 200\, \rm \mu Jy/beam$ for the GMRT. The resolution of the KAT-7 image is 160.10$\times$144.99 arcsec. The resolution of the  high resolution GMRT image is 7.44$\times$6.06 arcsec  while the resolution of the  tapered GMRT image is $44.89 \times 33.70\, \rm arcsec$.}
 \label{fig:Kat-7_GMRT}
\end{figure}

Figures~\ref{figure:highres} and~\ref{figure:lowres} show diffuse radio emission detected in MACS J2243.3-0935 by the GMRT at 610 MHz. In the GMRT images, the diffuse emission detected by KAT-7 is resolved into four distinct regions labelled A to D in Figure~\ref{figure:lowres}. Table~\ref{table:diff_flux} lists the flux densities for each of these regions. Flux densities were measured from the naturally weighted uvtapered image from within the 3$\sigma$ contour level. Errors in flux measurements were calculated using the formula: 
\begin{equation} \label{eq:fluxerror}
  \sigma_{ S_{\rm 610}}=\sqrt{\left(\sigma_{\rm cal}S_{610}\right)^{2}+\left(\sigma_{\rm rms}\sqrt{N_{\rm beam}}\right)^2},
\end{equation}
where $\sigma_{\rm cal}$ is the uncertainty in the calibration of the flux-scale and $N_{\rm beam}$ is the number of independent beams in the source. $\sigma_{\rm cal}$ is taken to be 10$\%$ for the GMRT \citep{Chandra2004}. Figure~\ref{fig:Kat-7_GMRT} shows the high resolution GMRT image of MACS J2243.3-0935 used to subtract the point sources with contours of the GMRT tapered image and KAT-7 image overlaid. The flux measured from the high resolution GMRT image within the same region of the KAT-7 cluster emission at a $3\sigma$ level is approximately 63 mJy. Extrapolating this flux to 1826 MHz assuming a spectral index of $\alpha=0.7$, where $S_{\nu}\propto \nu^{-\alpha}$, gives a value of approximately 29 mJy. Subtracting this from the KAT-7 flux calculated in \S~\ref{sec:KAT7_results} leaves a residual flux of 11 mJy.  The total flux measured from the point source subtracted, tapered GMRT image within the same region of the KAT-7 cluster emission is approximately 40 mJy. This suggests that the average spectral index of the diffuse emission in MACS J2243.3-0935 is 1.1.
\begin{table*}
 \centering
  \begin{threeparttable}[b]
  \caption{Properties of the diffuse emission in regions A, B, C and D. Column 1 is the region name, column 2 is the integrated flux density at 610 MHz, column 3 is the $k$-corrected integrated flux density as 610 MHz, column 4 is the surface brightness and column 5 is the $k$-corrected power at 610 MHz.}
  \label{table:diff_flux}
  \begin{tabular}{@{}lllll@{}}
  \hline
   Region & S$_{\rm 610MHz}$ (mJy) & S$_{\rm 610MHz,k-corr}$ (mJy) & I$_{\rm 610MHz}$ ($\mu$Jy/arcsec$^{2}$) & P$_{\rm 610MHz}$ (10$^{24}$ W Hz$^{-1}$)\\
 \hline
 A & $10.0\pm2.0$ & $12.0\pm2.0$ & $6.0\pm1.0$ &  $9.0\pm2.0$ \\
 B & $19.0\pm3.0$ & - & $11.0\pm2.0$ & -\\
 C & $11.0\pm2.0$ & - & $6.3\pm0.9$ & - \\
 D & $5.2\pm0.8$ & $4.7\pm0.7$ & $3.0\pm0.4$ & $3.4\pm0.5$\\
\hline
\end{tabular}
\begin{tablenotes}
\item [] $k$-corrected flux and $P_{\rm 601MHz}$ are calculated assuming a spectral index of 0.7 for region D and 1.4 for region A
\end{tablenotes}
\end{threeparttable}
\end{table*}

\subsubsection{Field Sources}
\begin{figure}
 \centering
   \includegraphics[width=0.5\textwidth]{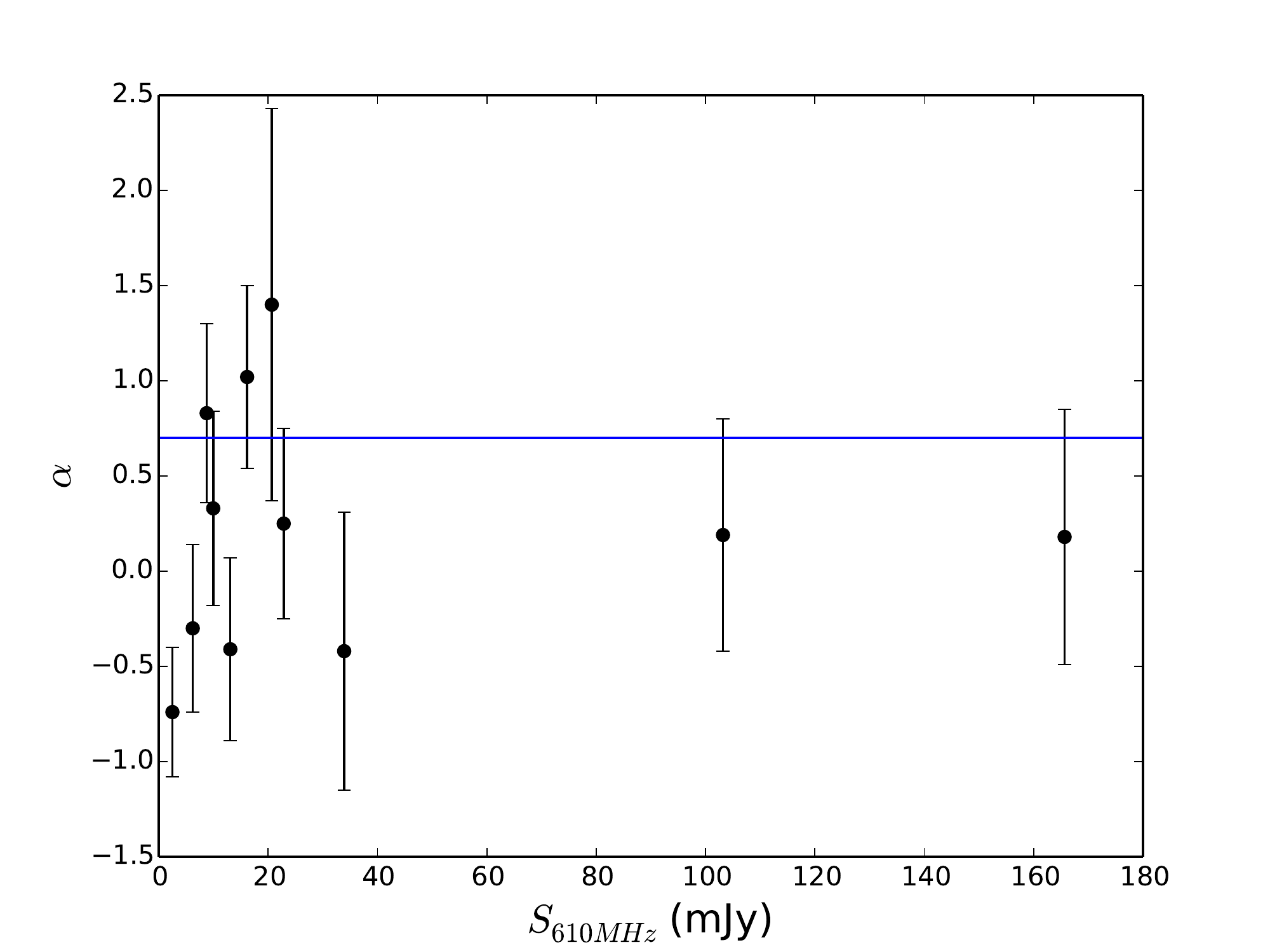}
  \caption{Spectral index of sources versus their flux density within the primary beam half power point. The blue line marks a spectral index of 0.7.  }
  \label{figure:alpha_dist}
\end{figure}

In order to further examine the fluxscale, the GMRT data were imaged with full uvrange, a Briggs weighting robust parameter of 0 and tapered close to the NVSS resolution. The software PYBDSM\footnote{PYBDSM documentation: \url{http://www.astron.nl/citt/pybdsm/}} was used to detect sources in both the NVSS and the GMRT maps. PYBDSM works by detecting all pixels in the map above a set peak threshold. It will then form islands of contiguous emission down to a set island threshold around the identified peak pixels. Gaussians are then fit to the islands and the Gaussians are grouped into individual sources. The flux densities of the sources are calculated by summing the flux densities of the Gaussians and the error in the flux density is calculated by summing the uncertainties in the gaussians in quadrature. For both the GMRT and the NVSS images, the peak threshold was set to 7$\sigma$ and the island threshold was set to 5$\sigma$. Table~\ref{table:field_sources} lists the detected sources and their flux densities in the GMRT and the NVSS maps. Table~\ref{table:field_sources} also shows the calculated spectral index of each source that is detected in both maps. Figure~\ref{figure:alpha_dist} shows the spectral indices for sources in the field versus their flux densities. The average spectral index is $0.2\pm0.6$. An average spectral index of $\alpha=0.7$ is expected for most optically thin extragalactic radio sources due to the energy distribution of cosmic rays produced in shocks. However at low frequencies, the spectral index distribution of faint sources is expected to have a flat tail due to the flattening of blazar spectra at frequencies below 1 GHz. \citep{Massaro2014}. 

\subsubsection{Optical Counterparts in Regions A-D}
\begin{figure*}
  \centering
  \begin{tabular}{cc}
     \subfloat[][]{\label{fig:A_high}}{\includegraphics[width=0.45\textwidth]{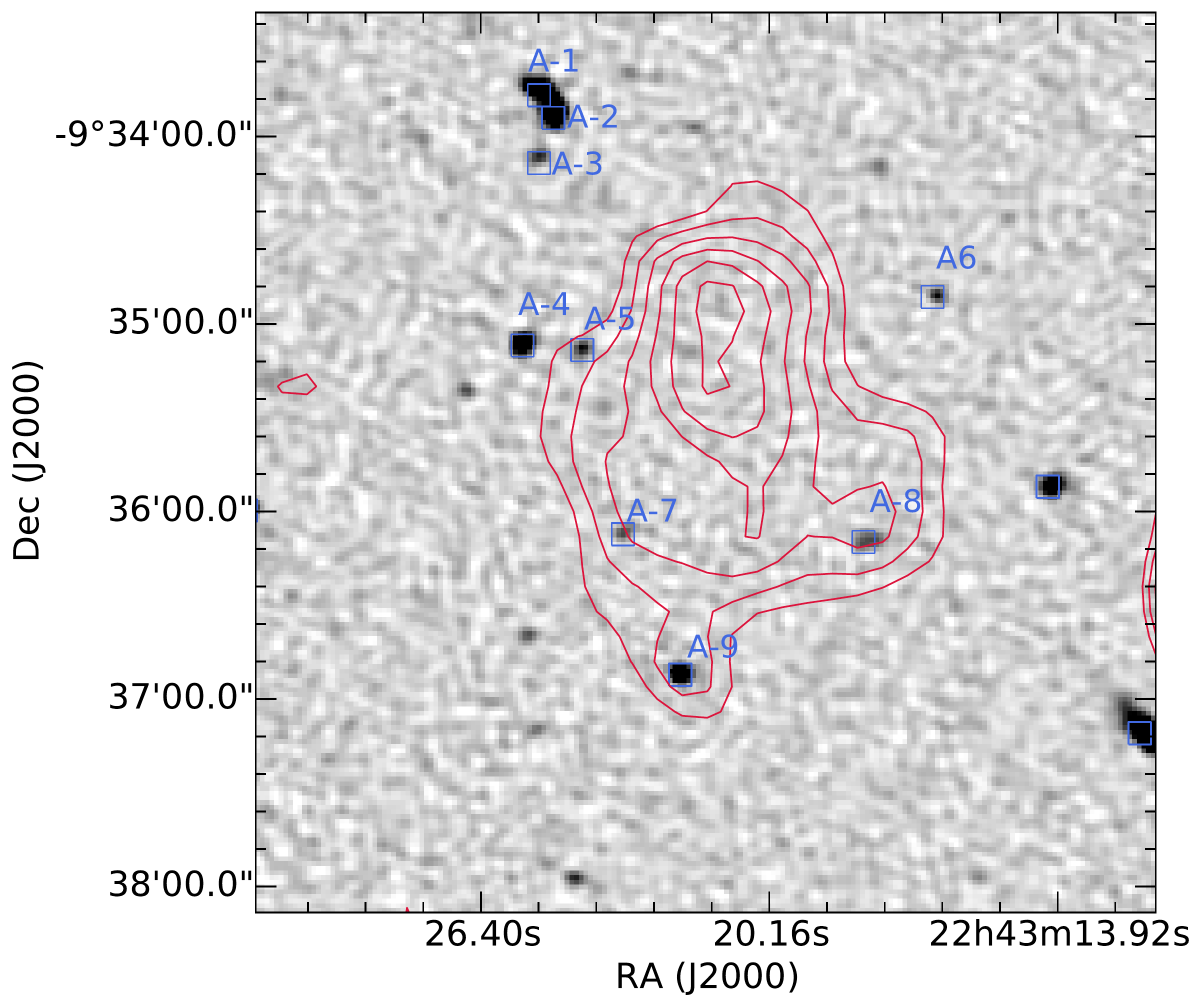}} & \subfloat[][]{\label{fig:A_low}}{\includegraphics[width=0.45\textwidth]{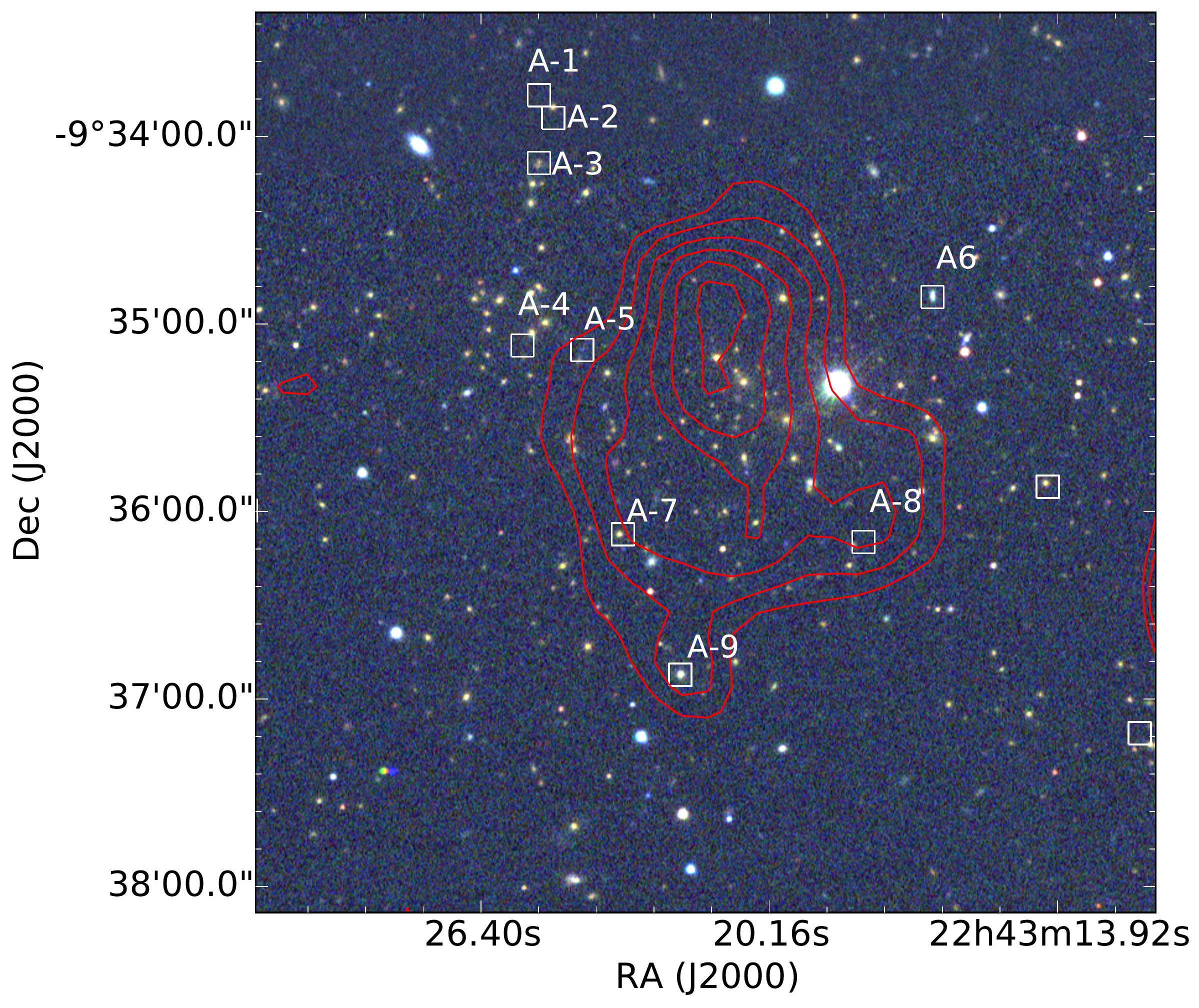}}\\ \subfloat[][]{\label{fig:B_high}}{\includegraphics[width=0.45\textwidth]{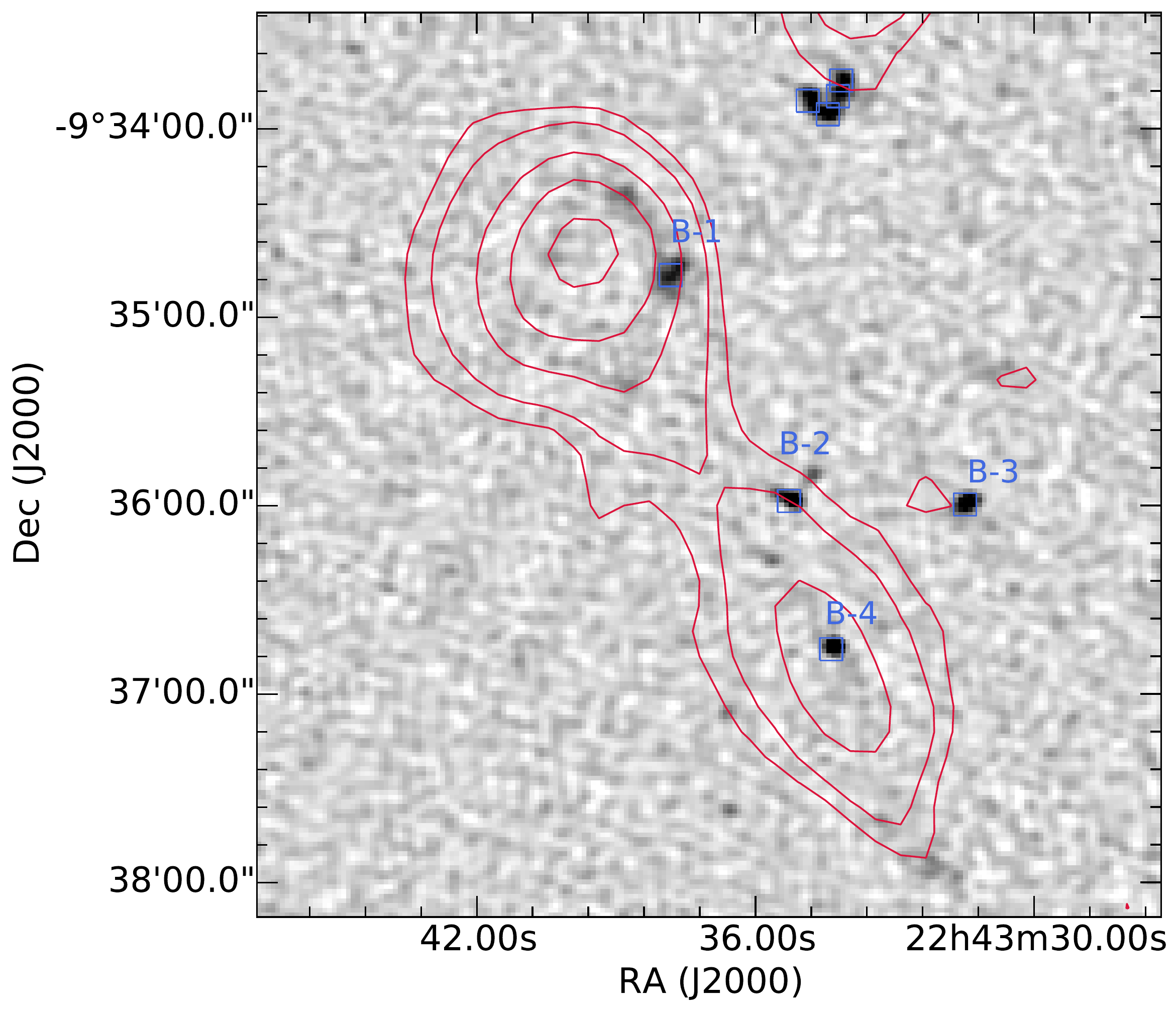}} & \subfloat[][]{\label{fig:B_low}}{\includegraphics[width=0.45\textwidth]{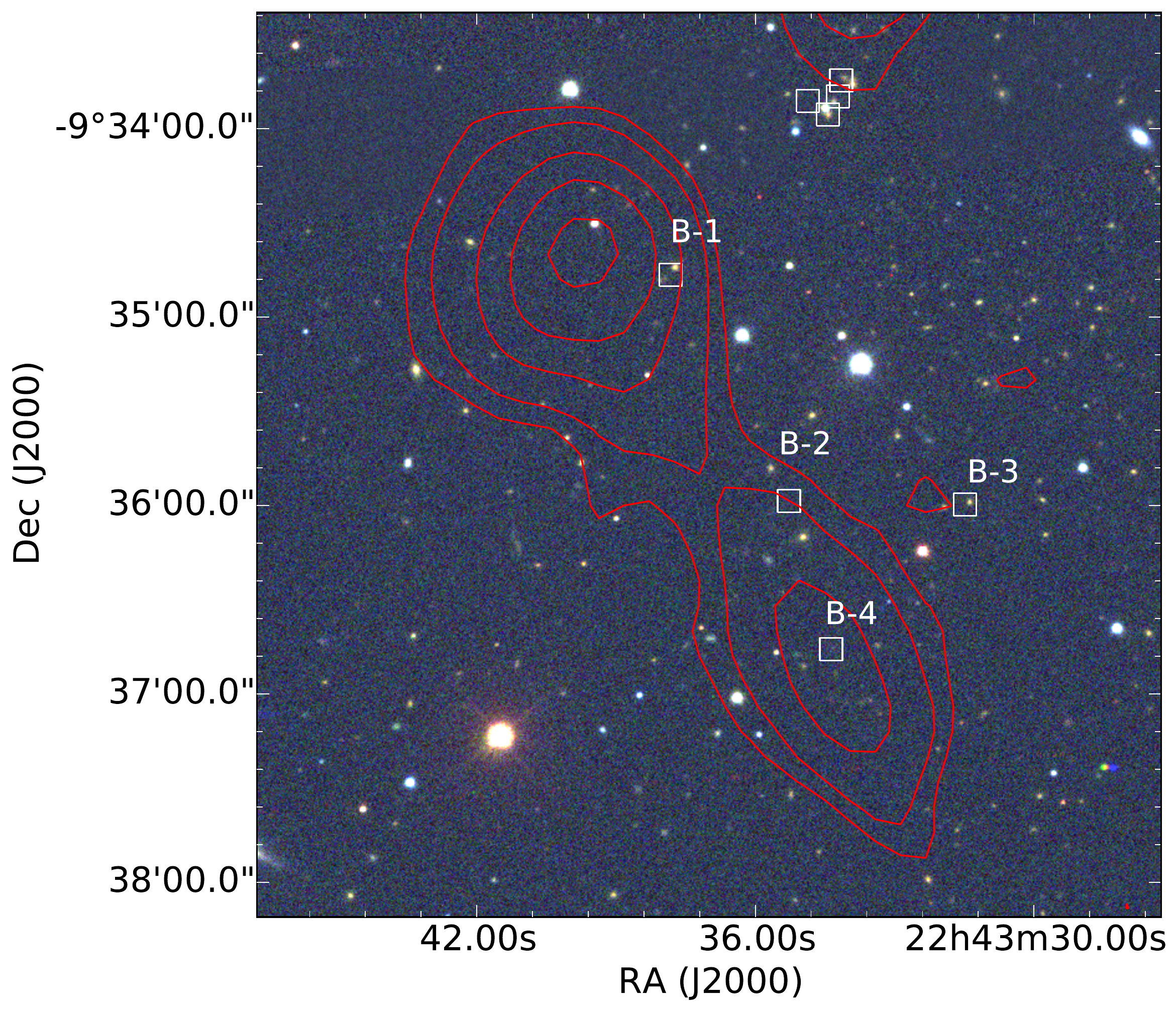}}\\ \subfloat[][]{\label{fig:C_high}}{\includegraphics[width=0.45\textwidth]{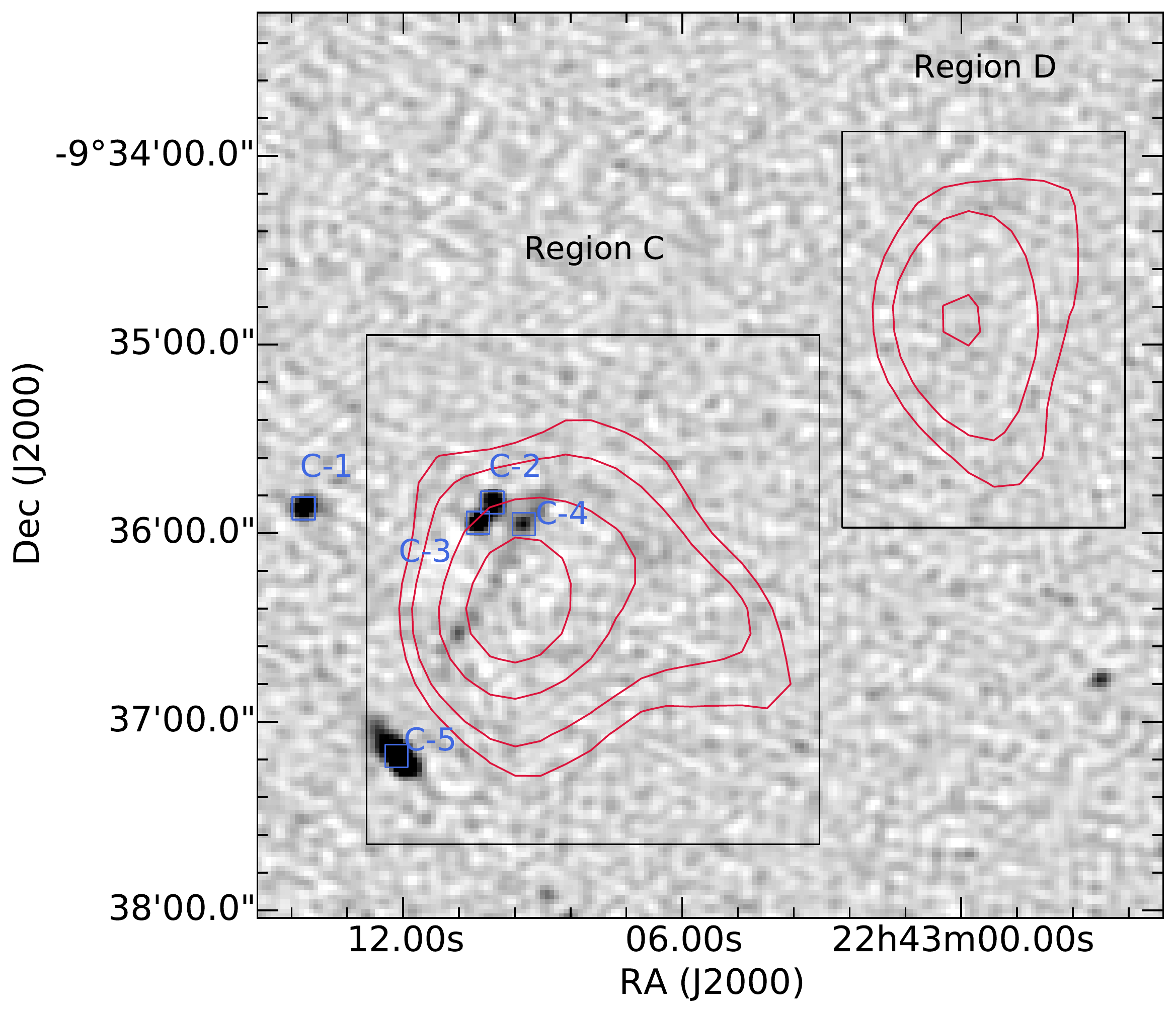}} &  \subfloat[][]{\label{fig:_C_low}}{\includegraphics[width=0.45\textwidth]{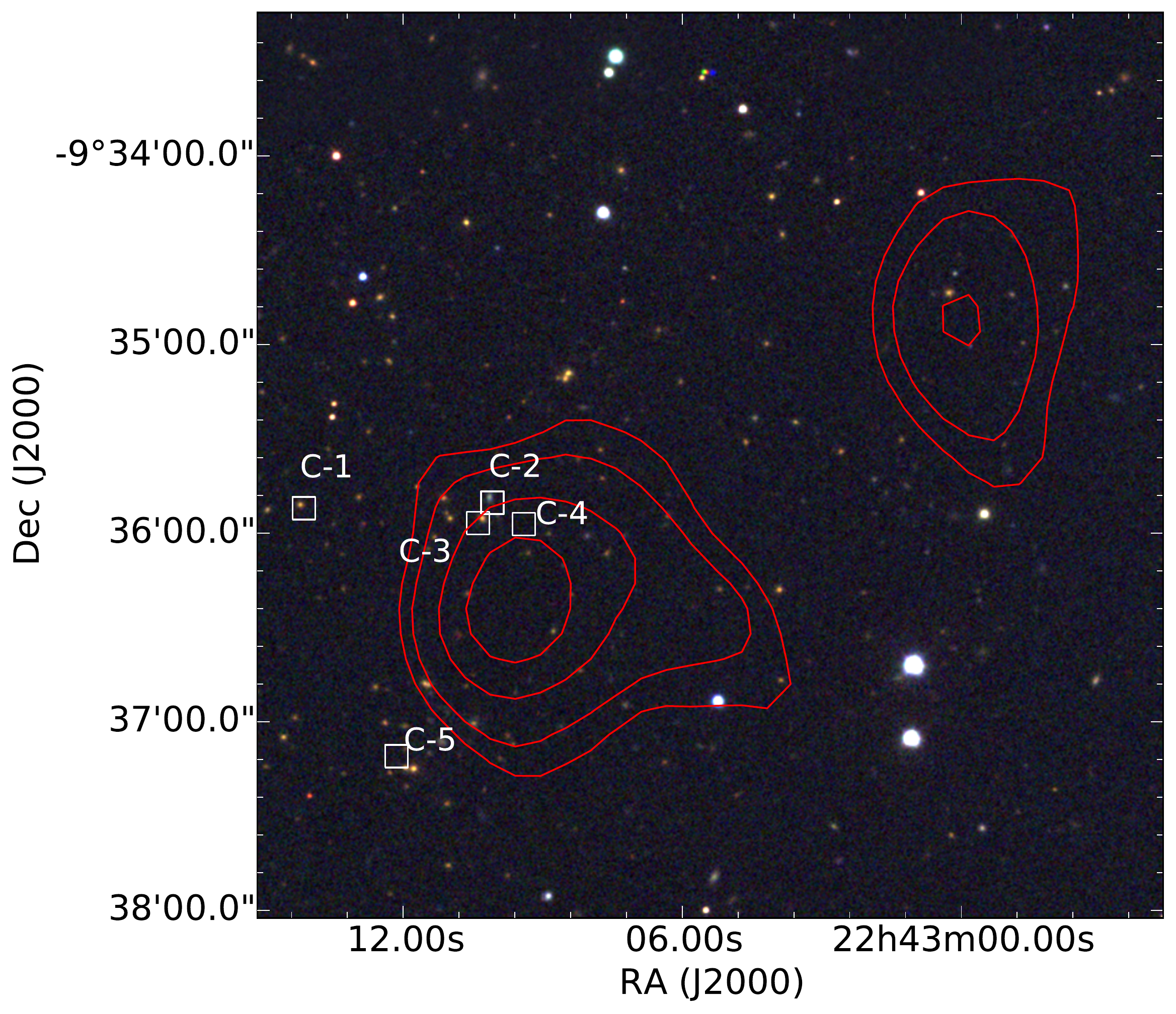}}\\
\end{tabular}
     \caption{Greyscale images show the robust 0 high resolution GMRT data in the left column and the rgb image of SDSS D12 i, r and g filters in the right column. The high resolution GMRT data has a rms noise of $40\, \rm \mu Jy/beam$ and a resolution of $4.84 \times 4.15\, \rm arcsec$. The uvtapered GMRT images are overlaid in each image. The resolution of the uvtapered GMRT image is $44.89\times 33.70 \, \rm arcsec$. Radio point sources have been subtracted from the uvtapered image using models extracted from the high resolution GMRT image shown in the greyscale. The locations of discrete radio sources detected by PYBDSM are marked by blue boxes in the left column and by white boxes in the right column. The first row shows images for region A with contours are at -3, 3, 4, 5, 6, 7, 8, 9, 10, 15, 20 $\times \sigma_{\rm rms}$ where $\sigma_{\rm rms} = 200\, \rm \mu Jy/beam$. The middle row shows images for region B and the last row shows images for regions C and D with contours at -3, 3, 5, 10, 15, 20 $\times \sigma_{\rm rms}$ where $\sigma_{\rm rms} = 200\, \rm \mu Jy/beam$.}
     \label{figure:Regions}
\end{figure*}

Figure~\ref{figure:Regions} shows the robust 0 images of regions A, B, C and D as well as the optical SDSS images of each region. The locations of discrete radio sources are marked. Galaxies within the redshift slice $z\pm0.04\left(1+z\right)$ are deemed to be associated with the cluster \citep{Wen2009}. Table~\ref{table:region_sources} lists the radio sources found in each region and their optical counterparts. An optical source was deemed to be the radio source's counterpart if the optical source was within one FWHM of the GMRT beam from the radio source. If more than one optical source lies with a FWHM of the radio source, then the source closest to the centroid of the radio source is deemed to be the optical counterpart.
\begin{table*}
 \centering
  \caption{Sources within GMRT primary beam half power point. Column 1-4 list the RA and Dec values of each source as well as error in the positions. Column 5 lists the integrated flux density at 610 MHz, column 6 is the integrated flux density measured from NVSS, column 7 is the offset between the source position measured in GMRT and NVSS. Column 8 is the spectral index of the source.}
  \label{table:field_sources}
  \begin{tabular}{@{}lllllccc@{}}
  \hline
R.A. (h:m:s) & R.A. err. (s) & Dec. (d:m:s) & Dec. err. (arcsec) & $S_{\rm 610,GMRT}$ (mJy) & $S_{\rm NVSS}$ (mJy) & Offset (arcsec) & $\alpha$\\
\hline
22:44:31.20 & 0.10 & -09:41:07.80 & 2.80 & 2.18$\pm$0.32 & - & - & -\\
22:44:26.90 & 0.00 & -09:26:36.40 & 0.30 & 33.91$\pm$0.73 & 48.16$\pm$1.22 & 3.1 & -0.42$\pm$0.09\\
22:44:29.60 & 0.10 & -09:45:39.50 & 1.30 & 6.22$\pm$0.44 & 6.37$\pm$0.75 & 1.3 & -0.03$\pm$0.38\\
22:44:12.80 & 0.00 & -09:34:22.60 & 0.50 & 13.09$\pm$0.48 & 18.34$\pm$0.99 & 1.4 & -0.41$\pm$0.18\\
22:44:12.60 & 0.10 & -09:46:29.40 & 2.30 & 2.50$\pm$0.34 & 4.63$\pm$0.73 & 5.6 & -0.74$\pm$0.58\\
22:44:02.40 & 0.00 & -09:51:18.40 & 0.10 & 165.66$\pm$0.67 & 143.02$\pm$1.32 & 19.2 & 0.18$\pm$0.03\\
22:44:03.20 & 0.10 & -09:28:32.70 & 0.90 & 9.96$\pm$0.51 & 7.56$\pm$0.82 & 6.0 & 0.33$\pm$0.33\\
22:43:56.60 & 0.10 & -09:25:26.00 & 1.60 & 3.79$\pm$0.37 & - & - & -\\
22:43:49.40 & 0.00 & -09:54:06.20 & 0.20 & 103.19$\pm$0.61 & 87.99$\pm$1.16 & 1.5 & 0.19$\pm$0.04\\
22:43:40.30 & 0.10 & -09:42:31.70 & 1.40 & 6.06$\pm$0.44 & - & - & -\\
22:43:39.10 & 0.10 & -09:34:42.30 & 1.50 & 10.41$\pm$0.49 & - & - & -\\
22:43:38.40 & 0.20 & -09:44:28.40 & 3.30 & 2.85$\pm$0.37 & - & - & -\\
22:43:35.60 & 0.10 & -09:30:42.60 & 2.40 & 2.92$\pm$0.36 & - & - & -\\
22:43:34.40 & 0.10 & -09:33:41.30 & 2.90 & 3.69$\pm$0.35 & - & - & -\\
22:43:28.60 & 0.10 & -09:28:00.50 & 1.30 & 7.62$\pm$0.44 & - & - & -\\
22:43:27.80 & 0.10 & -09:47:10.00 & 2.50 & 1.91$\pm$0.29 & - & - & -\\
22:43:25.80 & 0.00 & -09:39:11.40 & 0.50 & 16.16$\pm$0.48 & 6.91$\pm$0.90 & 3.2 & 1.02$\pm$0.37\\
22:43:24.90 & 0.00 & -09:33:49.50 & 0.90 & 20.67$\pm$1.03 & 6.45$\pm$0.77 & 2.3 & 1.40$\pm$0.36\\
22:43:23.70 & 0.10 & -09:50:52.60 & 2.80 & 3.10$\pm$0.36 & - & - & -\\
22:43:18.80 & 0.00 & -09:18:54.40 & 0.30 & 22.85$\pm$0.50 & 18.54$\pm$0.99 & 1.5 & 0.25$\pm$0.16\\
22:43:18.60 & 0.10 & -09:44:52.20 & 1.00 & 8.76$\pm$0.47 & 4.40$\pm$0.63 & 2.2 & 0.83$\pm$0.42\\
22:43:09.20 & 0.10 & -09:36:09.60 & 1.20 & 21.34$\pm$0.85 & - & - & -\\
22:43:00.00 & 00.20 & -09:34:56.20 & 3.90 & 2.89$\pm$0.36 & - & - & -\\
22:42:53.50 & 0.10 & -09:22:49.10 & 2.30 & 3.33$\pm$0.35 & - & - & -\\
22:42:16.30 & 0.10 & -09:31:53.10 & 2.40 & 2.49$\pm$0.35 & - & - & -\\

\hline
\end{tabular}
\end{table*}

\begin{table*}
 \centering
  \caption{Sources within different regions. Column 1 is the region name and column 2 is the source name. Column 3 and 4 are the position of the sources in RA and Dec. Column 5 is the integrated flux density at 610 MHz, column 6 is the name of the SDSS optical counterpart, column 7 is the offset between the position of the GMRT source and the SDSS counterpart. Column 8 is the sources position relative to the cluster. }
  \label{table:region_sources}
  \begin{tabular}{@{}lclllccc@{}}
  \hline
Region & Source & RA (h:m:s) & Dec (d:m:s) & S$_{610}$ (mJy) & SDSS source & Offset (arcsec) & Position\\
\hline
Region A & A-1 & 22:43:25.1 & -09:33:46.8 & 3.62$\pm$0.07 & SDSS J224324.84-093350.9  & 3.05 & cluster member\\
 & A-2 & 22:43:24.8 & -09:33:54.0 & 5.18$\pm$0.07 & SDSS J224324.84-093350.9  & 3.05 & cluster member\\
 & A-3 & 22:43:25.1 & -09:34:08.5 & 0.61$\pm$0.07 & SDSS J224325.14-093408.7 & 0.25 & cluster member\\
 & A-4 & 22:43:25.5 & -09:35:06.8 & 1.94$\pm$0.07 & SDSS J224325.30-093503.1  & 4.68 & cluster member\\
 & A-5 & 22:43:24.2 & -09:35:08.3 & 0.52$\pm$0.08 & - & - & -\\
 & A-6 & 22:43:16.6 & -09:34:51.3 & 0.48$\pm$0.08 & SDSS J224316.63-093451.4 & 0.17 & foreground source\\
 & A-7 & 22:43:23.3 & -09:36:07.2 & 0.37$\pm$0.08 & SDSS J224323.40-093607.5 & 1.17 & cluster member\\
 & A-8 & 22:43:18.1 & -09:36:09.8 & 0.84$\pm$0.06 & - & - & -\\
 & A-9 & 22:43:22.1 & -09:36:52.2 & 2.02$\pm$0.08 & SDSS J224322.08-093652.4 & 0.18 & foreground source\\
\hline
Region B & B-1 & 22:43:37.8 & -09:34:46.6 & 2.73$\pm$0.05 & SDSS J224337.71-093444.5 & 2.76 & cluster member\\
 & B-2 & 22:43:35.3 & -09:35:58.7 & 1.47$\pm$0.07 & - & - & -\\
 & B-3 & 22:43:31.5 & -09:35:59.7 & 1.46$\pm$0.07 & SDSS J224331.40-093558.9 & 1.47 & cluster member\\
 & B-4 & 22:43:34.4 & -09:36:45.6 & 1.24$\pm$0.08 & - & - & -\\
\hline
Region C & C-1 & 22:43:14.1 & -09:35:52.1 & 1.71$\pm$0.07 & SDSS J224314.19-093551.1 & 1.12 & cluster member\\
 & C-2 & 22:43:10.1 & -09:35:50.3 & 1.56$\pm$0.07 & SDSS J224310.12-093548.7 & 1.74 & foreground source\\
 & C-3 & 22:43:10.4 & -09:35:56.8 & 1.50$\pm$0.07 & SDSS J224310.28-093555.4 & 2.09 & cluster member\\
 & C-4 & 22:43:09.4 & -09:35:57.1 & 1.31$\pm$0.06 & - & - & -\\
 & C-5 & 22:43:12.1 & -09:37:10.8 & 6.83$\pm$0.08 & SDSS J224311.92-093714.8 & 5.15 & cluster member\\
 \hline
\end{tabular}
\end{table*}

\subsubsection{Region A}  
\label{sec:regA_results}
Region A is detected in the tapered image shown in Figure~\ref{figure:lowres} at a 5$\sigma$ level. It is not detected at a significant level in either the untapered image, shown in Figure~\ref{figure:highres}, or the high resolution image used to subtract the point sources. The emission at the 3$\sigma$ level fills roughly the same region as the X-ray emission, shown in Figure~\ref{figure:lowres}. The radio emission appears to be extended along an axis almost perpendicular to the extension of the X-ray emission. In Figure~\ref{fig:A_high} there are no compact radio sources coincident with or near the peak of region A and so the diffuse emission is unlikely to be associated with a single discrete source. At a redshift 0.447, region A has a largest linear scale (LLS) of approximately 0.92 Mpc.

\subsubsection{Region B}
To the east of the cluster, complex diffuse emission can be seen in both the tapered and untapered image. The emission has a LLS of approximately 1.7 Mpc. In Figure~\ref{fig:B_high} there are two peaks in the emission. The southern peak is coincident with a discrete radio source, B-4, at J22:43:34.4 -09:35:58.6. There is no SDSS, X-ray or infra-red counterpart for this source. It is possible that the optical source  has a high redshift that puts it outside the range of SDSS. This would place the source behind the cluster.
\begin{figure}
  \includegraphics[width=0.5\textwidth]{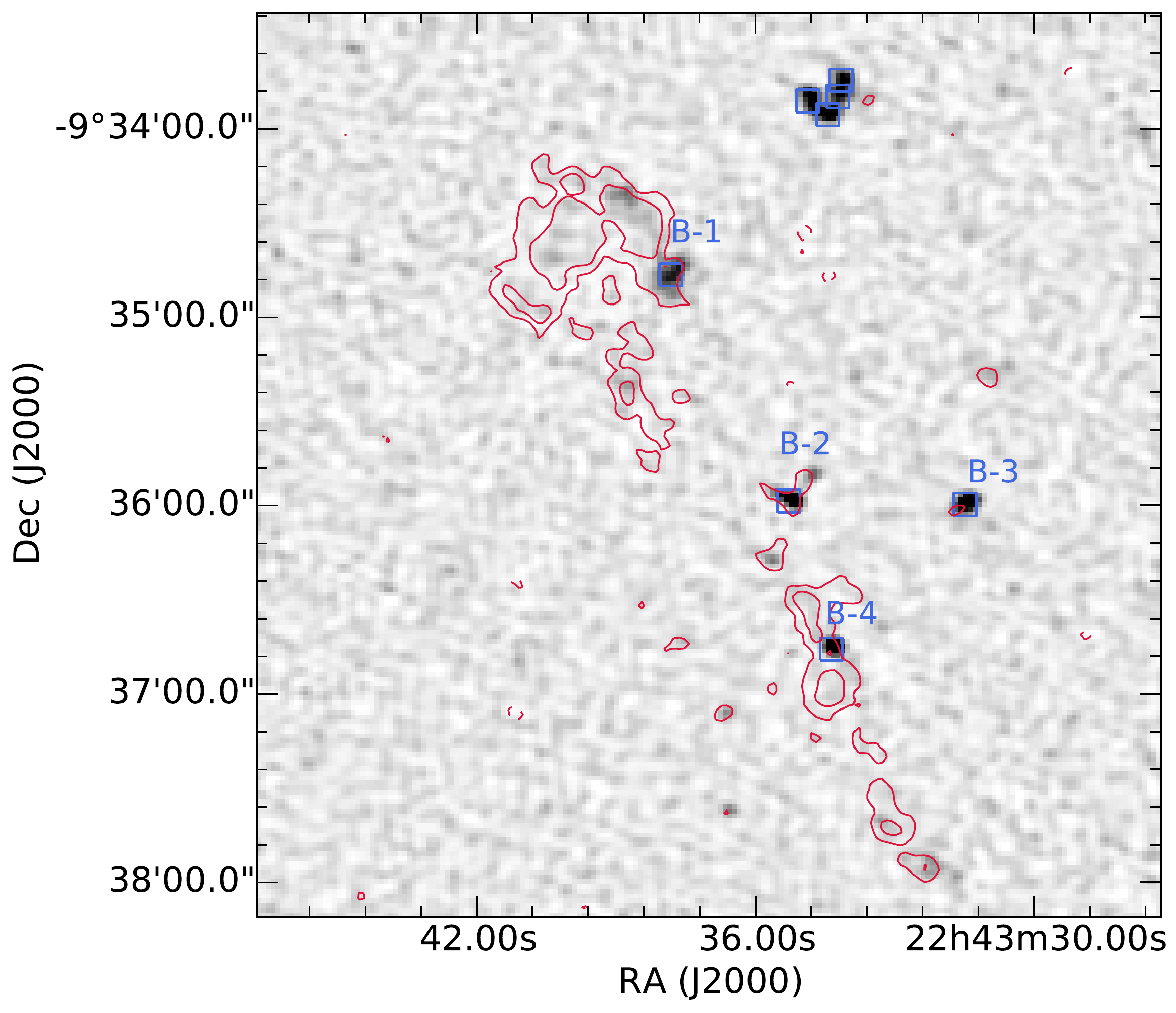}
  \caption{Greyscale images show the robust 0 high resolution GMRT data for region B. Red contours show naturally weighted image. Contours are at -3, 3, 5, 10, 15, 20 $\times$ $\sigma_{\rm rms}$ where $\sigma_{\rm rms}=45\, \rm \mu Jy/beam$. The resolution of the GMRT image is 7.44$\times$6.06 arcsec. Radio point sources have been subtracted from the naturally weighted image using models extracted from the high resolution GMRT image shown in the greyscale. The blue squares mark discrete radio sources.}
  \label{figure:Region_B_natural}
\end{figure}
The northern peak is centred near source B-1 at J22:43:37.8 -9:34:46, which is located within the cluster. In Figure~\ref{figure:Region_B_natural} the northern peak is resolved into an arc of emission with one end of the arc coincident with B-1. In the untapered image there is no emission detected connecting the northern and southern areas of region B.

\subsubsection{Regions C and D}
\begin{figure}
  \includegraphics[width=0.5\textwidth]{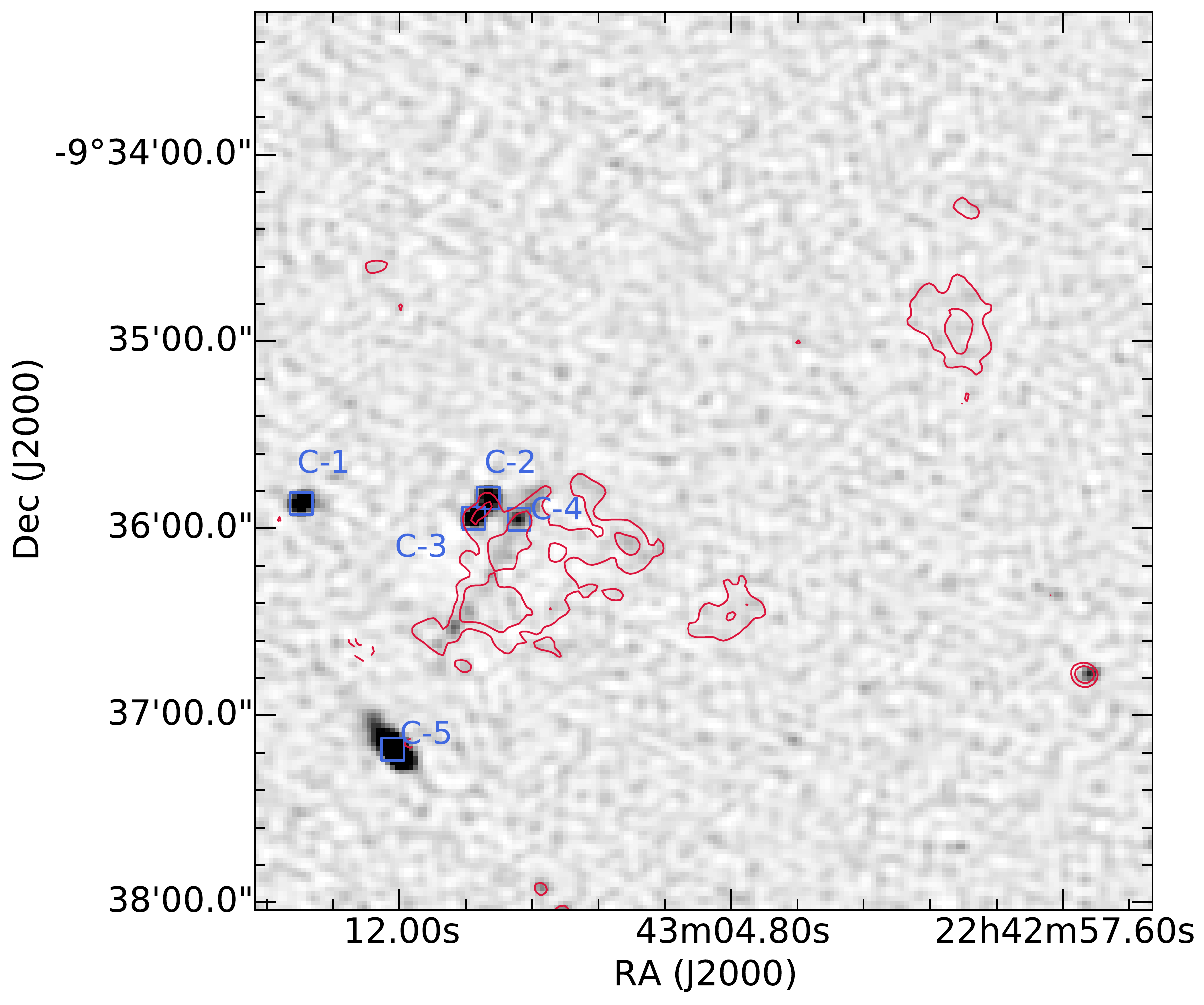}
  \caption{Greyscale images show the robust 0 high resolution GMRT data for regions C and D. Red contours show naturally weighted image. Contours are at -3, 3, 5, 10, 15, 20 $\times$ $\sigma_{\rm rms}$ where $\sigma_{\rm rms}=45\, \rm \mu Jy/beam$. The resolution of the GMRT image is 7.44$\times$6.06 arcsec. Radio point sources have been subtracted from the naturally weighted image using models extracted from the high resolution GMRT image shown in the greyscale. The blue squares mark discrete radio sources.}
  \label{figure:Region_C_nat}
\end{figure}

To the west of the cluster, there is a second region of complex diffuse emission. Again this can be seen in both the tapered and untapered maps. There appear to be two separate sources. The first, region C, is brighter and appears to coincident with at least two  discrete sources, C-2 and C-3. A narrow linear structure is evident in the highest resolution greyscale image as well as the untapered image of region C in Figure~\ref{figure:Region_C_nat}. The linear structure extends from the north-west to south-east. Region C has a LLS of approximately 0.76 Mpc

The second region, region D, is on the edge of the cluster's virial radius. There are no discrete radio or optical sources evident in the region. It has a LLS of approximately 0.68 Mpc. The eastern side of region D is curved while the western side of region D is somewhat flatter. 

\section{Discussion}
\label{sec:disscusion}

\subsection{A Giant Radio Halo in MACS J2243.3-0935}

Figure~\ref{figure:lowres} shows the X-ray emission of the cluster with the radio contours of region A overlaid. The morphology, size and position of region A are consistent with that of a giant halo.

\subsubsection{Spectral Index}
\label{sec:spectral_index}
As discussed in \S~\ref{sec:regA_results}, Region A is clearly detected in the GMRT 610 MHz image, however it is not detected in the NVSS image and in the KAT-7 image all cluster emission is unresolved. Without high resolution data at 1.822 GHz, it is not possible to disentangle emission in region A from emission in region B, C or D in the KAT-7 image. The NVSS image does not have the resolution or the sensitivity to subtract the discrete sources from the KAT-7 image. As such we are only able to put a lower limit on the spectral index of the radio halo using the NVSS image. The rms noise of NVSS is 0.45 mJy/beam. Thus a 3$\sigma$ upper limit flux density for the radio halo at 1400 MHz is 1.35 mJy/beam. Assuming the halo has the same spatial extent at 1400 MHz this gives an integrated upper limit on integrated flux of 8.2 mJy. Taken with the 610 MHz flux density of $10.0\pm2.0$ mJy this gives a lower limit on the spectral index of $\alpha \geq 0.28$. Radio Halos are expected to have much steeper spectral indices than 0.28, however NVSS does not have the surface brightness sensitivity to more tightly constrain the spectral index of region A.

\citet{Feretti2012} suggest a link between the average temperature of a cluster and the spectral index of radio halos. They find that radio halos in clusters with an average temperature between 8 and 10 keV have an average spectral index of $\alpha=1.4\pm 0.4$. MACS J2243.3-0935 has an temperature of $8.24\pm0.92$ K and so an estimate spectral index of $\alpha=1.4$ will be used in this paper to estimate the properties of the halo in MACS J2243.3. 

\subsubsection{Scaling Relations}

\begin{figure*}
 \centering
\subfloat[][]{\includegraphics[width=0.5\textwidth]{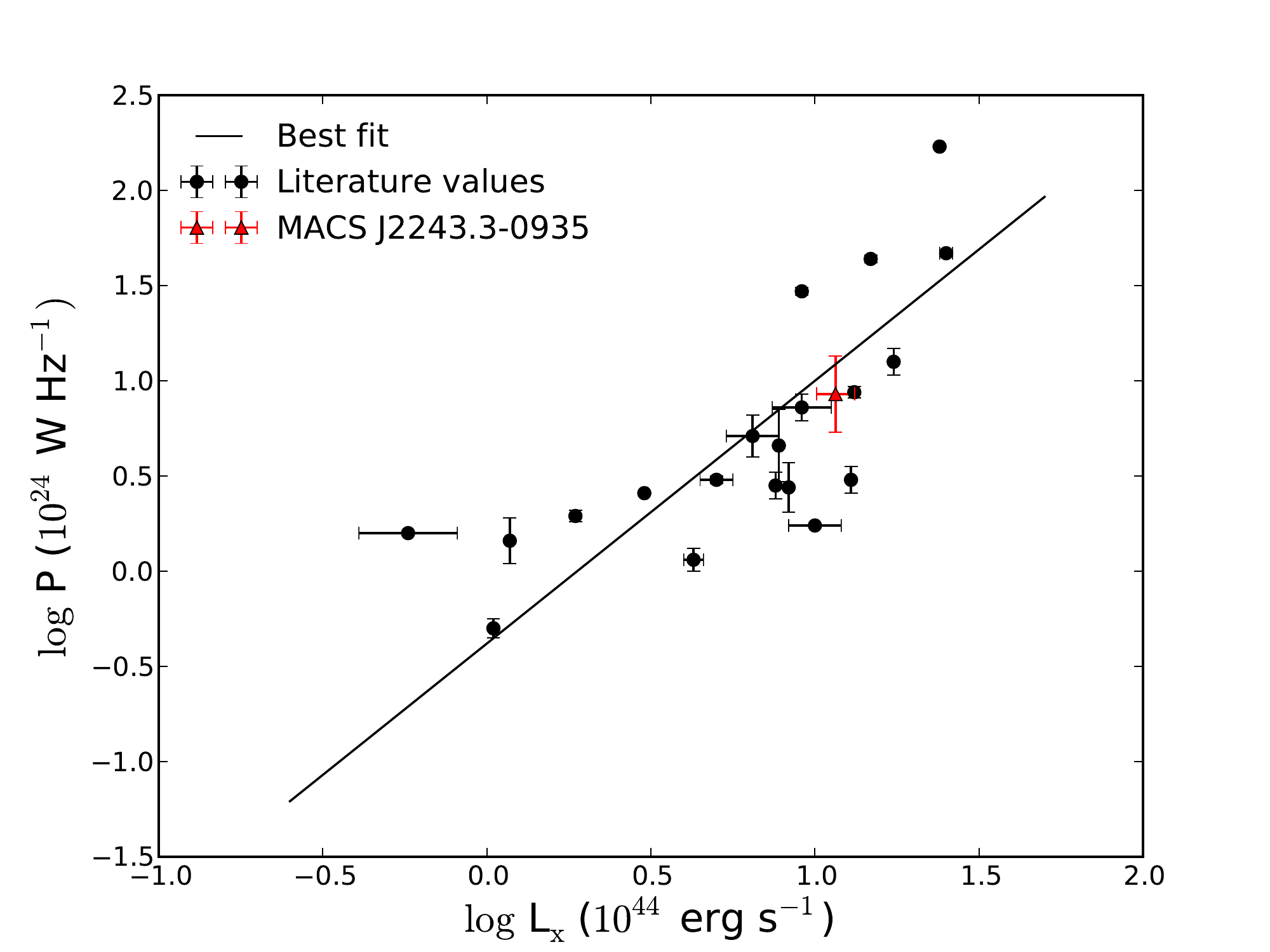} } \subfloat[][]{\includegraphics[width=0.5\textwidth]{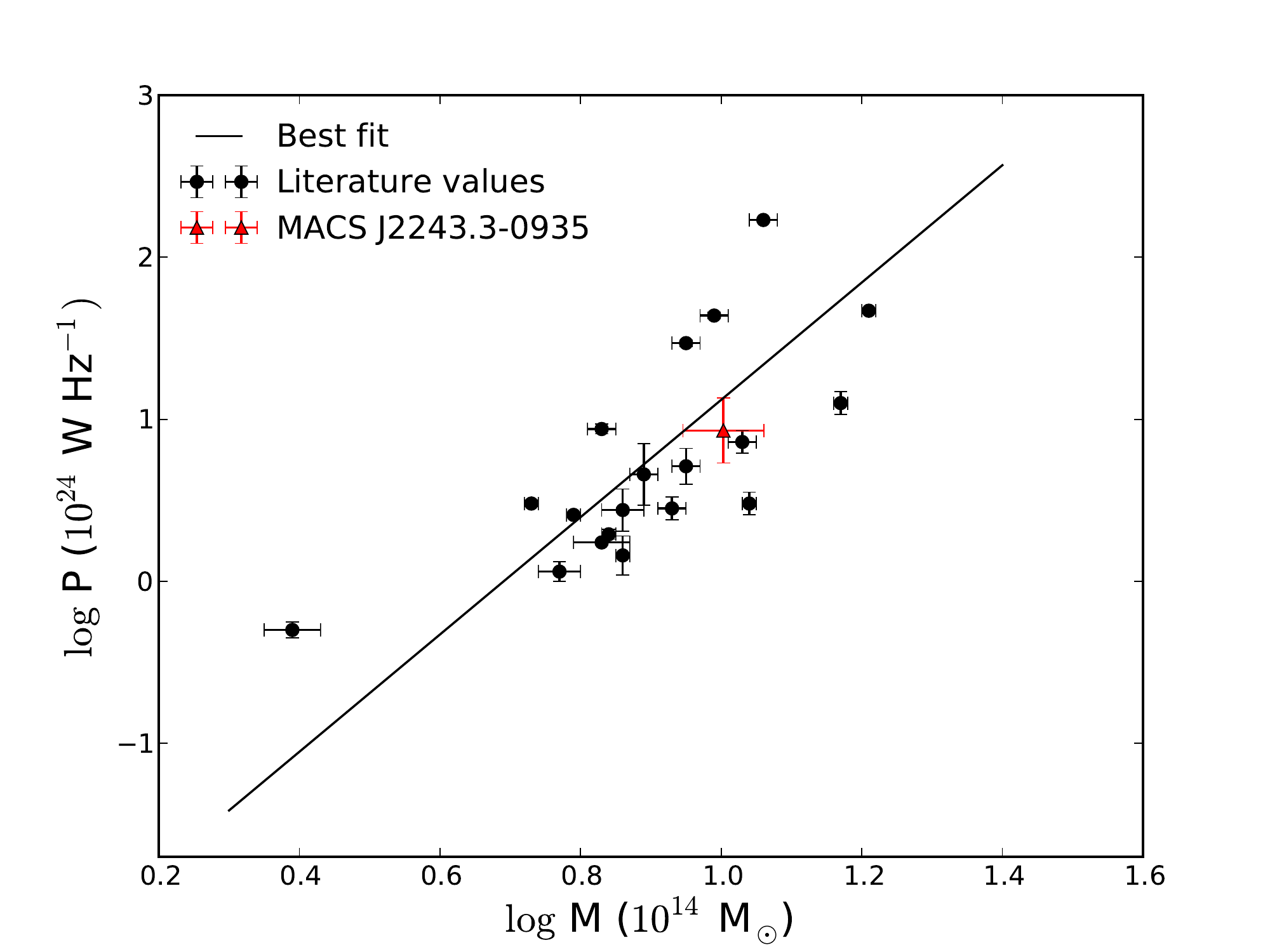}}
     \caption{These plots marks the position of the halo in MACS J2243.3-0925 in red on the 610 MHz scaling relations examined in \citet{Yuan2015} (a) shows the $P_{610\, \rm MHz}-L_{\rm x}$ correlation and (b) shows the $P_{610\, \rm MHz}-M$ relation. Black data points are taken from \citet{Yuan2015}.  }
    \label{figure:scaling relations}
\end{figure*}

Using the spectral index stated above, the $k$-corrected radio power of the halo at 610 MHz is $\left(9.0\pm2.0\right)\times10^{24}$ W Hz$^{-1}$. Figure~\ref{figure:scaling relations} shows the halo's position on the $P_{\rm 610\, \rm MHz}-L_{\rm x}$ and $P_{\rm 610\, \rm MHz}-M_{\rm 500}$ diagrams. Figure~\ref{figure:scaling relations} is a reproduction of Figure 2 in \citet{Yuan2015} with the data point for MACS J2243.3-0935 included. Region A in MACS J2243.3-0935 is in good agreement with the power expected from the correlations show in Figure~\ref{figure:scaling relations}, providing further evidence that region A is a radio halo. 

\subsubsection{Equipartition B-Fields}
\label{sec:equiB}

\citet{Beck2005} provide a revised formula for the classical equipartition magnetic field,
\begin{equation}  \label{eq:equiB}
  B=\left\{\frac{4\pi\left(2\alpha+1\right)\left(K_{0}+1\right)I_{\nu}E_{p}^{1-2\alpha}\left(\frac{\nu}{2c_{1}}\right)^{\alpha}}{\left(2\alpha-1\right)c_{2}lc_{4}}\right\}^{\frac{1}{\alpha+3}},
\end{equation}
where $\alpha$ is the spectral index, $K_{0}$ is the ratio of proton energy densities to electron energy densities, $I_{\nu}$ is the synchrotron intensity, $E_{\rm p}$ is the proton rest energy, $\nu$ is the observing frequency and $l$ is the extent of the source along the line of sight. $c_{1}$ and $c_{3}$ are constants while $c_{2}$ is a function of the spectral index and $c_{4}$ is a function of the inclination of the source. See Appendix A in \citet{Beck2005} for definition of these variables.

There is much discussion in the literature on the precise value of $K_{0}$. Different CR injection mechanisms predict different values for $K_{0}$ for the ICM. For example, turbulent acceleration predicts $K_{0}=100$, production of secondary CRe predicts $K_0$ in the range of 100 to 300 and first order Fermi shock acceleration predicts values of $K_0$ in the range of 40 to 100 \citep{Beck2005}. However energy losses such as synchrotron and inverse Compton could inflate the value of $K_0$ to values much greater than 100. \citet{Vazza2014} compare some of the current models for CR injection to radio and Fermi data on clusters. They find that values of $K_{0}\geq100$ require gamma emission above the derived Fermi upper limits, suggesting that $K_{0}\leq100$.

\begin{figure}
  \includegraphics[width=0.5\textwidth]{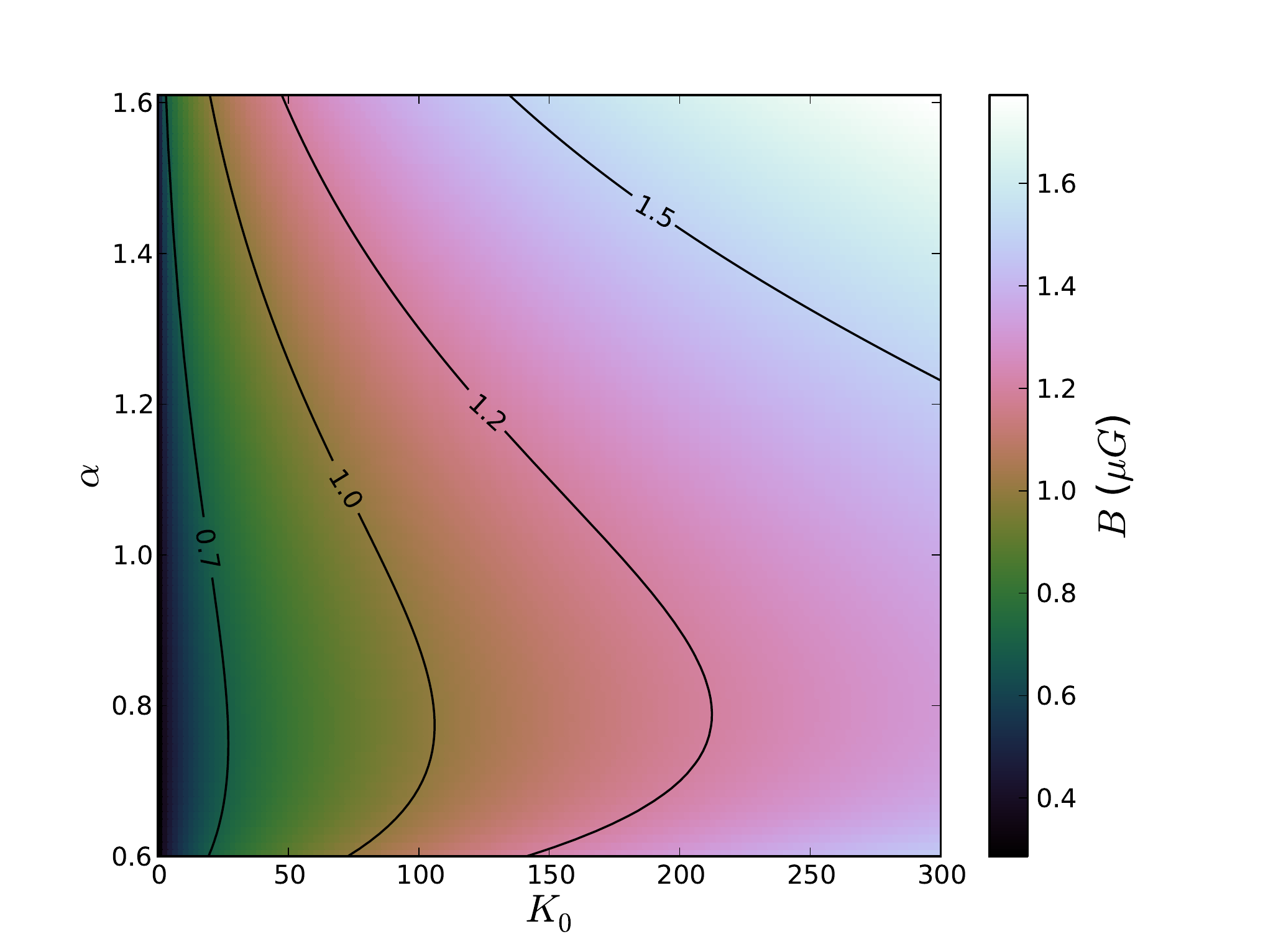}
  \caption{Equipartition magnetic field strength for region A for a range of values of $\alpha$ and $K_{0}$. Black contours mark regions of constant magnetic field strength. }
  \label{figure:equipartion_A}
\end{figure}
\begin{figure}
 \includegraphics[width=0.5\textwidth]{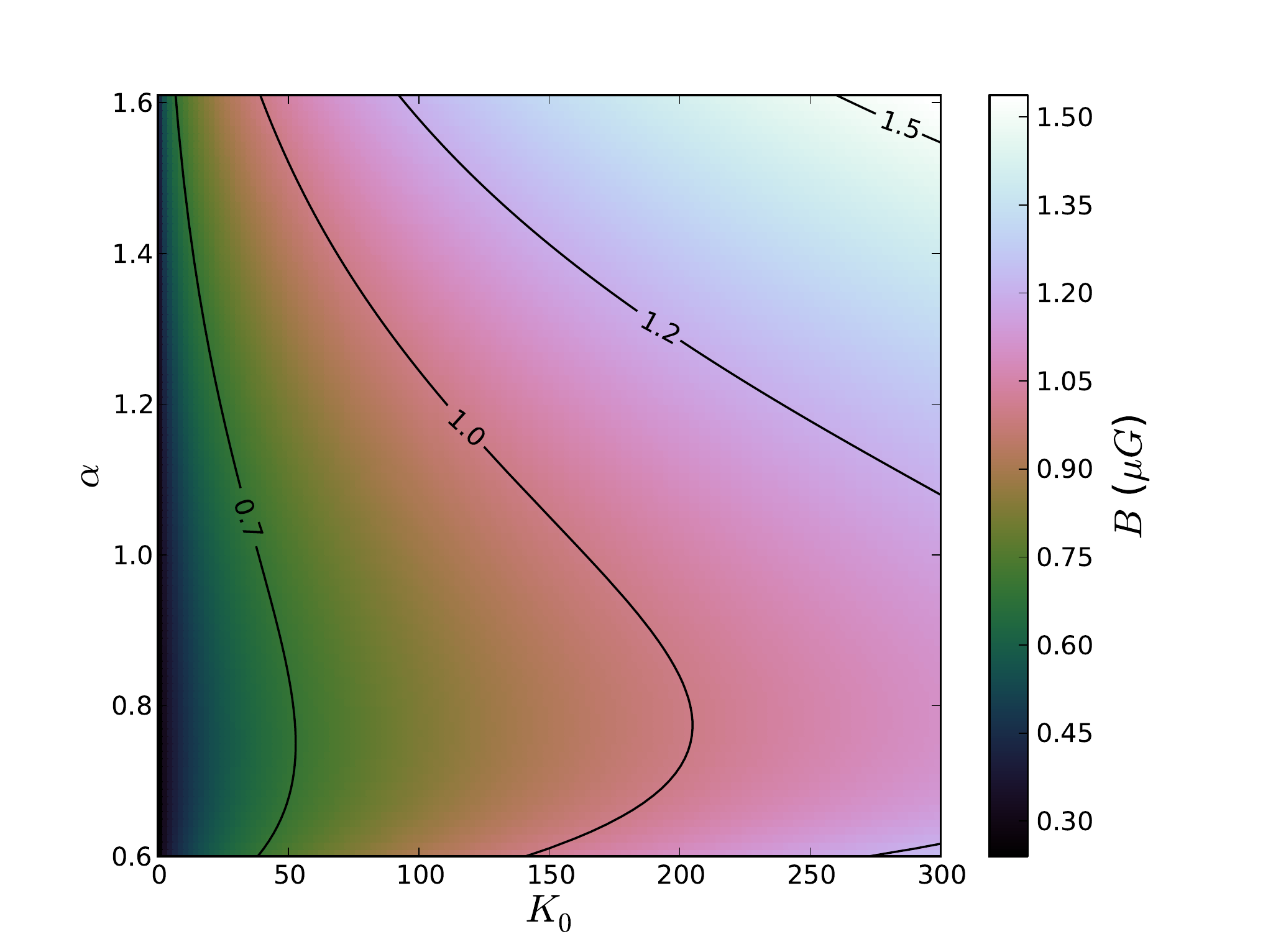}
 \caption{Equipartition magnetic field strength for region D for a range of values of $\alpha$ and $K_{0}$. Black contours mark regions of constant magnetic field strength. }
 \label{figure:equipartion_D}
\end{figure}

Using Equation \ref{eq:equiB}, we calculate the equipartition magnetic field of the cluster in region A and region B for different values of $\alpha$ and $K_{0}$. Figure~\ref{figure:equipartion_A} shows the results for region A while Figure~\ref{figure:equipartion_D} shows the results for region B. For both regions, magnetic field strengths vary from less than 0.5 $\mu$G for flat spectral indices and small values of $K_{0}$ to $1.5\,\mu \rm{G}$ for steep spectral indices and high values of $K_{0}$.

\subsection{Possible Radio Relics in MACS J2243.3-0935}

\subsubsection{Region B}
There are four possible explanations for the diffuse emission in Region B. The first is that it is merely a superposition of emission from sources at different redshifts. The second is that the emission is associated with the interaction of sources at the same redshift. The third is that the emission is from a giant radio galaxy (GRG). And finally the emission could be a radio relic.

The LLS of region B is consistent with both a GRG or a radio relic. The position of region B at the periphery of a cluster is expected for a radio relic while GRGs are more commonly found in less dense regions such as galaxy groups \citep{Malarecki2015}. The double peaked morphology of region B is unusual for a radio relic. A possible explanation for these localised regions of increased emission, in the context of a radio relic, is that these peaks coincide with areas of fossil AGN activity. The fossil CRe in these areas would be accelerated to higher energies and lead to a localised increase in emission.

 The northern peak in figure~\ref{figure:Region_B_natural} could also be interpreted as a bow shock in the lobe of a GRG. However it seems more likely that this is instead emission associated with B-1 and that B-1 is a head-tail radio galaxy.  

 While the linear size of region B is consistent with a radio relic or GRG, comparison between figure~\ref{figure:highres} and figure~\ref{figure:lowres} suggest that region B is not a region of continuous emission, but instead a series of radio galaxies that are unresolved in the tapered image, thus ruling out a radio relic and GRG. This would place B-1 and B-4 as the sources of the northern and southern peaks respectively. 

B-4 and B-2 do not have optical counterparts in SDSS.  They are likely background galaxies outside the redshift range of SDSS, suggesting that region B is a superposition of sources at different redshifts.

\subsubsection{Region C}
Figure~\ref{figure:Region_C_nat} shows the high resolution GMRT data for region C with the naturally weighted contours overlaid. The morphology of region C seen in Figure~\ref{figure:Region_C_nat} is consistent with bent tail radio galaxy. Bent tail galaxies are commonly found in galaxy clusters where the dense ICM warps the radio jet \citep{Mao2009,Pratley2013}. Such sources have been used to probe different physical properties of the ICM, including the density \citep{Freeland2008} and magnetic field strength \citep{Clarke2001,Vogt2003,Pratley2013}.  However due the cluster of discrete radio sources, C-2, C-3 and C-4, a superposition of sources can't be ruled as an explanation for the emission in region C.

The  position  of region C in the cluster and its LLS are also consistent with that of a radio relic. There is evidence to suggest that some radio relics are generated when shocks re accelerated fossil plasma from radio galaxies \citep{Ensslin2001b,Bonafede2014,Shimwell2015}. In such a scenario, region B could be a produced when fossil emission from C-2, C-3 or C-4 was reaccelerated by a merger shock. However without knowing how the spectral index varies across the source it is not possible to differentiate between a radio relic and a bent tail radio galaxy. 


\subsubsection{Region D}


NVSS and FIRST do not detect any radio sources in Region D. The rms noise of NVSS is 0.45 mJy/beam. Thus a 3$\sigma$ upper limit flux density for region D at 1400 MHz is 1.35 mJy/beam. Assuming region D has the same spatial extent at 1400 MHz this gives an upper limit on the integrated flux density of 2.0 mJy. Taken with the 610 MHz flux density of 5.2mJy this gives a lower limit on the spectral index of $\alpha\geq0.7$.

The greyscale image in Figure~\ref{figure:Region_C_nat} shows data from baselines longer than 4 k$\lambda$ imaged with a Briggs robust parameter of 0 for both region C and region D with the contours for the naturally weighted, point source subtracted image overlaid. No compact radio sources can be seen in this image coincident with or near the peak of region D and so the diffuse emission is unlikely to be associated with a single discrete source.

With a LLS of 0.68 Mpc, Region D is consistent with that of a radio relic. The lack of radio point sources in region D suggests this emission is not associated with a discrete source. Region D is about twice the length on the north-south axis as on the east-west axis, which is consistent with the morphology of elongated radio relics. Radio relics are likely formed by shocks produced by either major/minor cluster mergers or through the infall of the warm-hot intergalactic medium (WHIM) onto the cluster. Shocks produced by cluster mergers are expected to have a Mach number less than 5 \citep{Skillman2008}. \citet{Hong2014} study the properties of shocks at cluster outskirts and suggest that around half of radio relics with Mach numbers greater than 3, as well as relatively flat radio spectra, are infall shocks. To date only a  few relics have been described as infall relics in the literature. For example, \citet{Brown2011} suggest that the radio relic 1253+275 in the Coma cluster is caused by the infalling group NGC 4839 while \citet{PfrommerJones2011} model the structure of the head  tail radio galaxy NGC 1265 by assuming that the galaxy passed through an accretion shock onto the Perseus cluster. \citet{PfrommerJones2011} calculate the Mach number of the inferred accretion shock in the Perseus cluster to be approximately $\mathcal{M}=4.2$.

\begin{figure}
 \centering
 \includegraphics[width=0.5\textwidth]{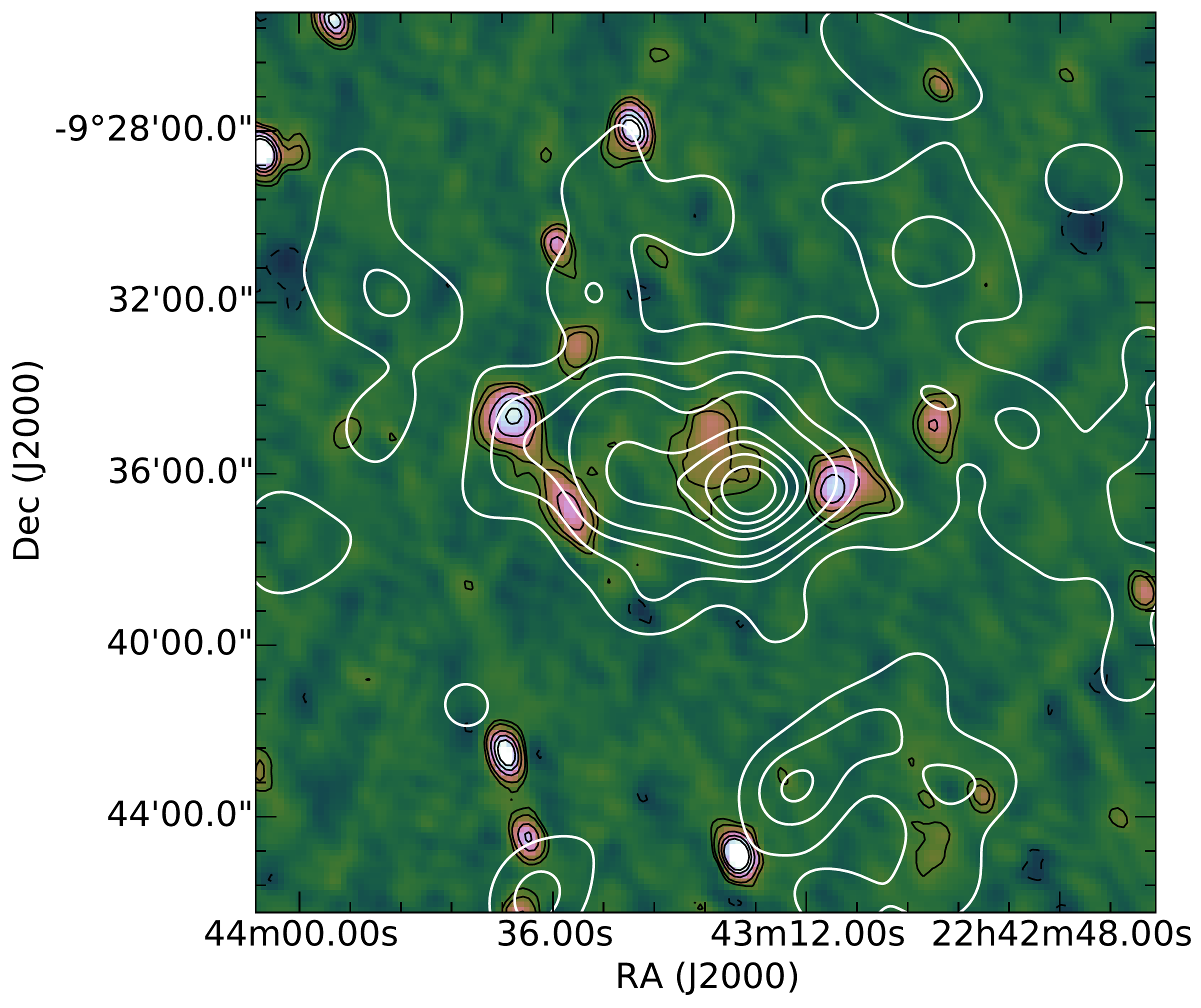}
 \caption{Colourscale image of diffuse emission in MACS J2243.3-0935 with black contours overlaid showing the tapered image. Contours are at -3, 3, 5, 10, 15, 20 $\times$ $\sigma_{\rm rms}$ where $\sigma_{\rm rms}=200\, \rm \mu Jy/beam$. Galaxy density contours overlaid in white. These contours are range from 20$\%$ to 90$\%$ of the peak value in steps of 10.
 \label{figure:galaxy_density}}
\end{figure}

MACS J2243.3-0935 is the central cluster in the supercluster SCL2243-0935. Figure~\ref{figure:galaxy_density} shows the number density of SDSS galaxies in the region of MACS J2243-0935. The galaxy density at each point was calculated by counting the number of galaxies that fell in each pixel.  The resultant map was then smoothed with a Gaussian of width 42 arcsec. Galaxies were chosen to be in the photometric redshift bin $0.39<z<0.5$ and within 4 Mpc of the cluster centre. This redshift bin was chosen to account for uncertainties in the photometric redshift error and a large spatial region was chosen to include the filamentary structure in the supercluster. Region D is located at the virial radius where one of the supercluster filaments, AH in \citet{Schirmer2011}, meets MACS J2243.3-0935. The location of region D as well as the LLS are suggestive of an infall relic, however multi-frequency analysis would be required to properly determine the nature of the region. 

\begin{figure*}
  \centering
  \begin{tabular}{cc}
     \subfloat[][]{\label{fig:IRAS12}}{\includegraphics[width=0.5\textwidth]{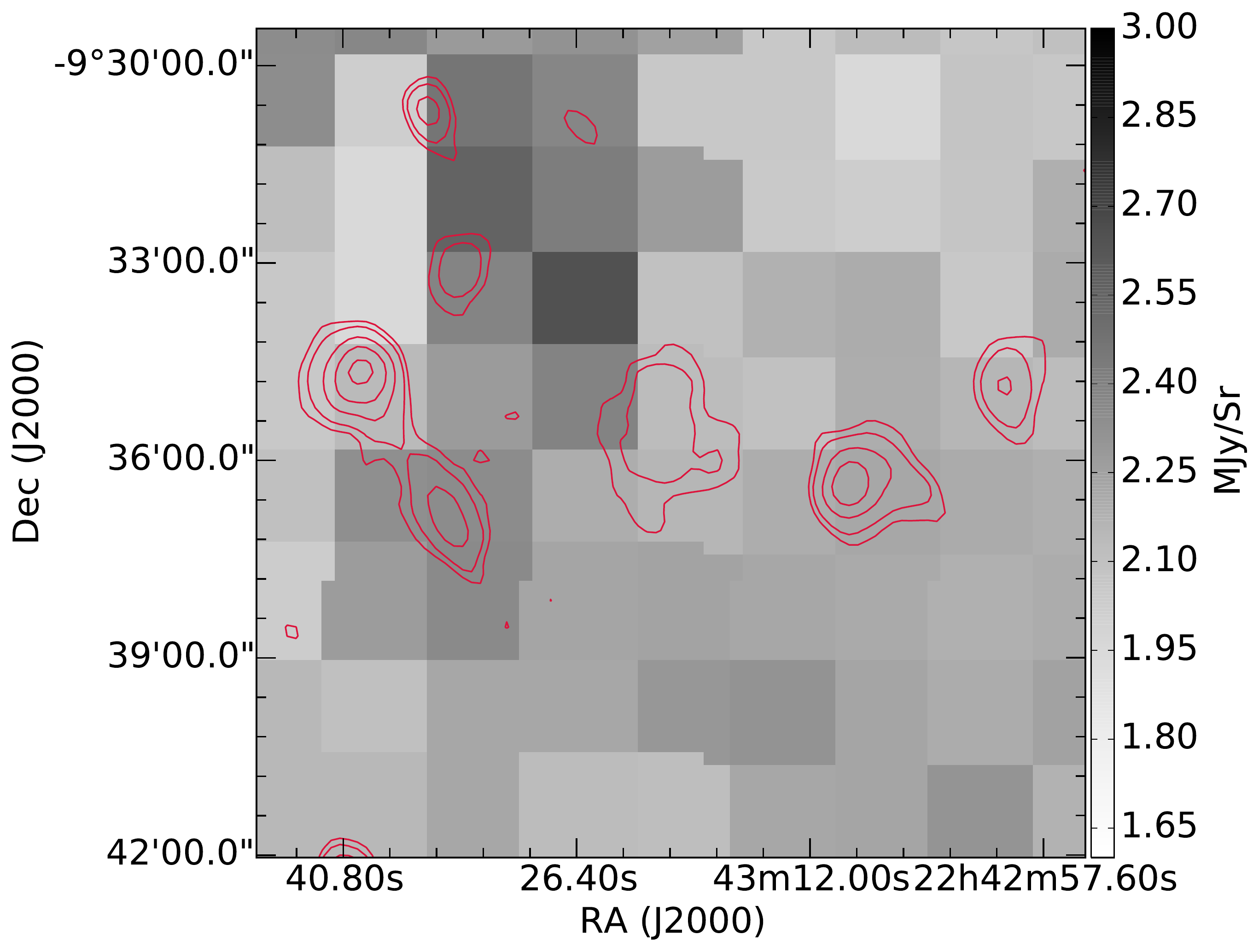}} & \subfloat[][]{\label{fig:IRAS25}}{\includegraphics[width=0.5\textwidth]{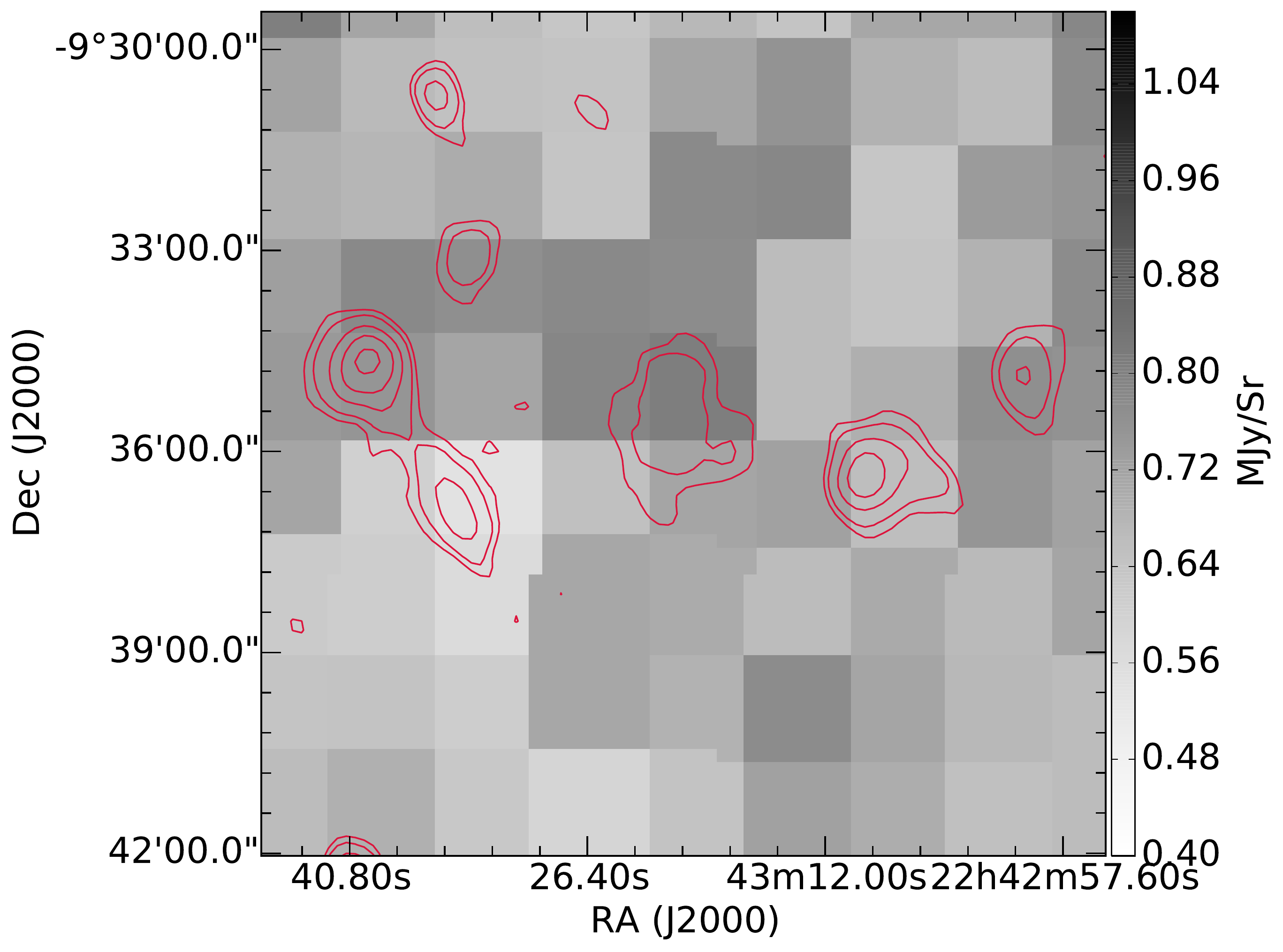}}\\ \subfloat[][]{\label{fig:IRAS60}}{\includegraphics[width=0.5\textwidth]{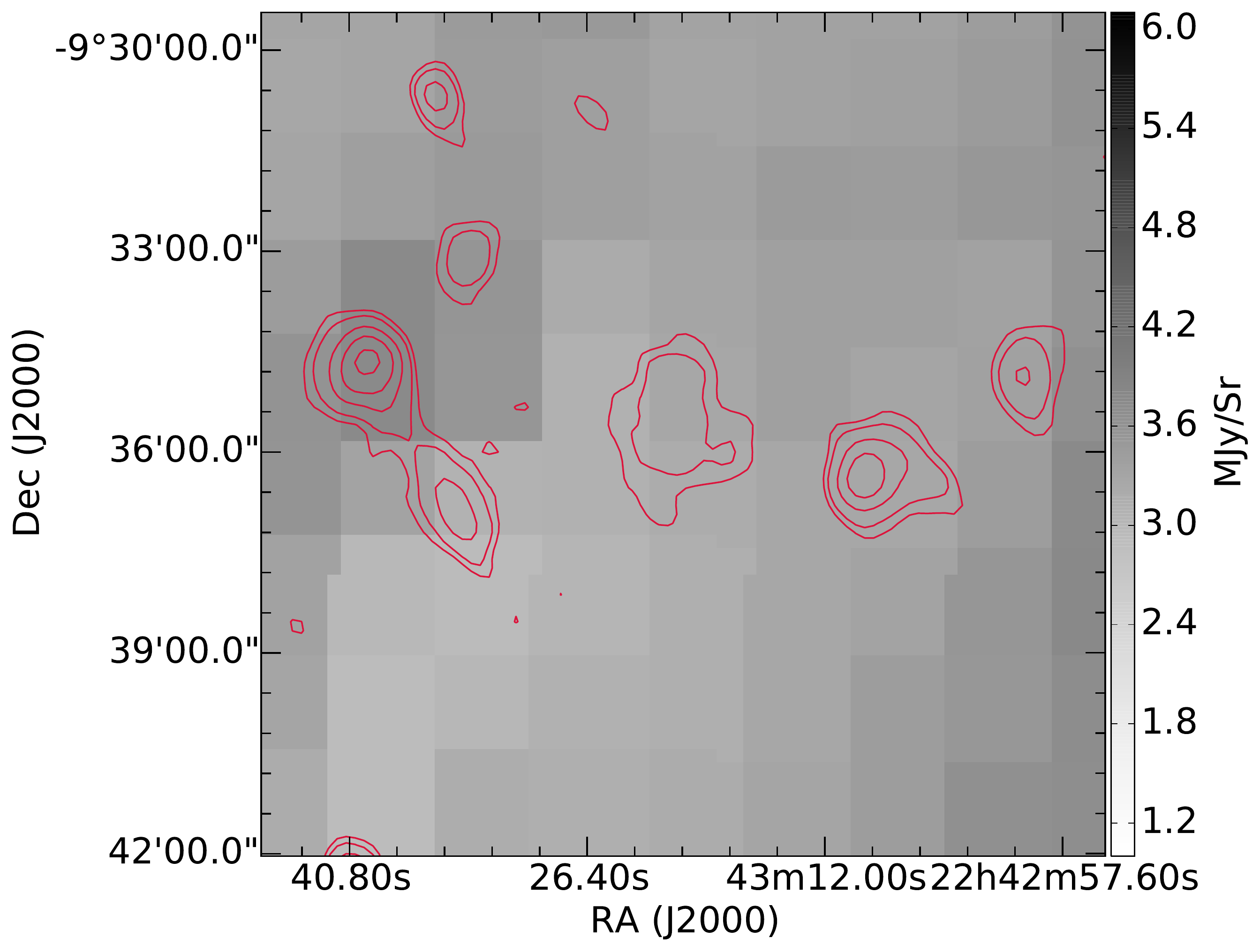}}&
\subfloat[][]{\label{fig:IRAS100}}{\includegraphics[width=0.5\textwidth]{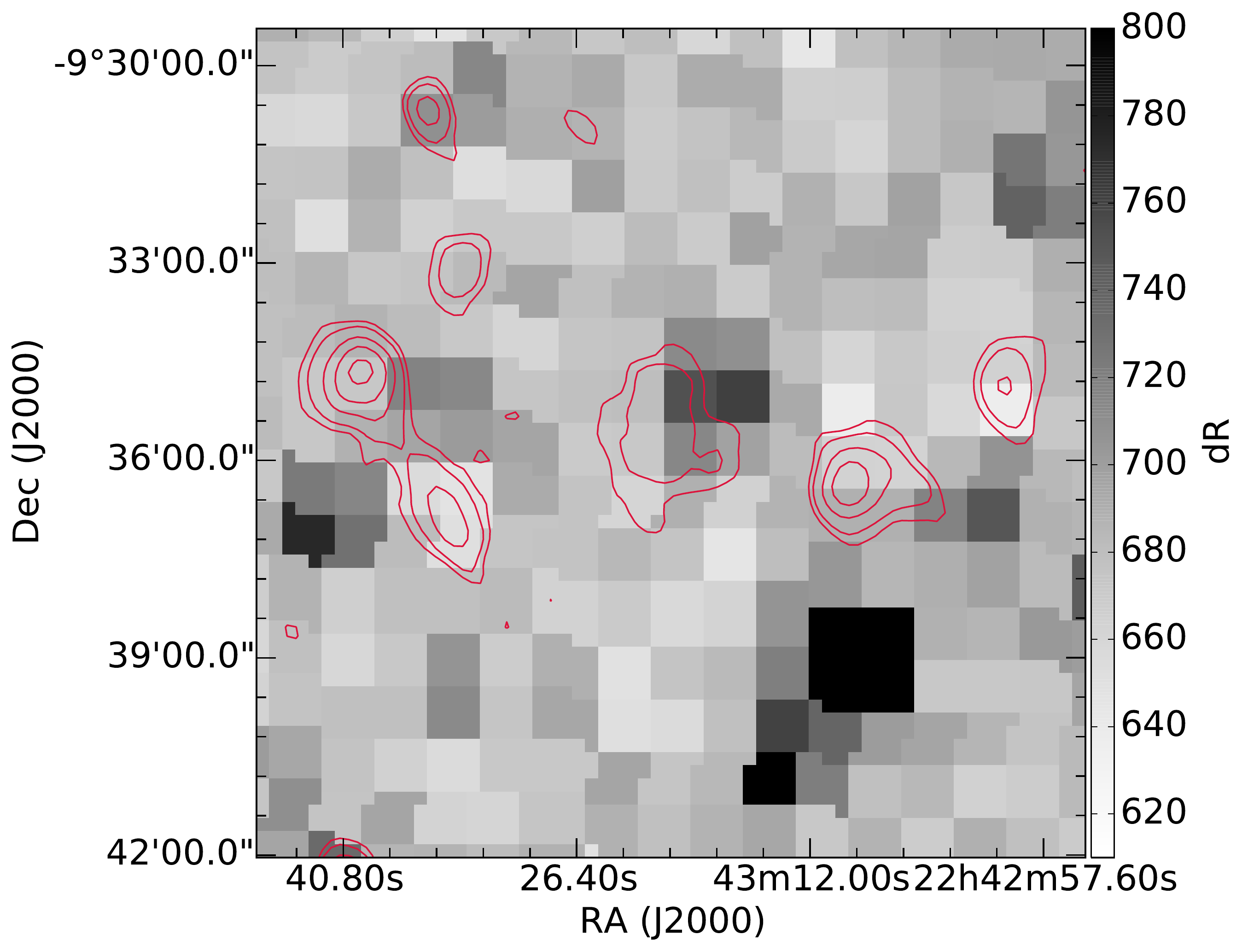}}\\ \subfloat[][]{\label{fig:SHASSH}}{\includegraphics[width=0.5\textwidth]{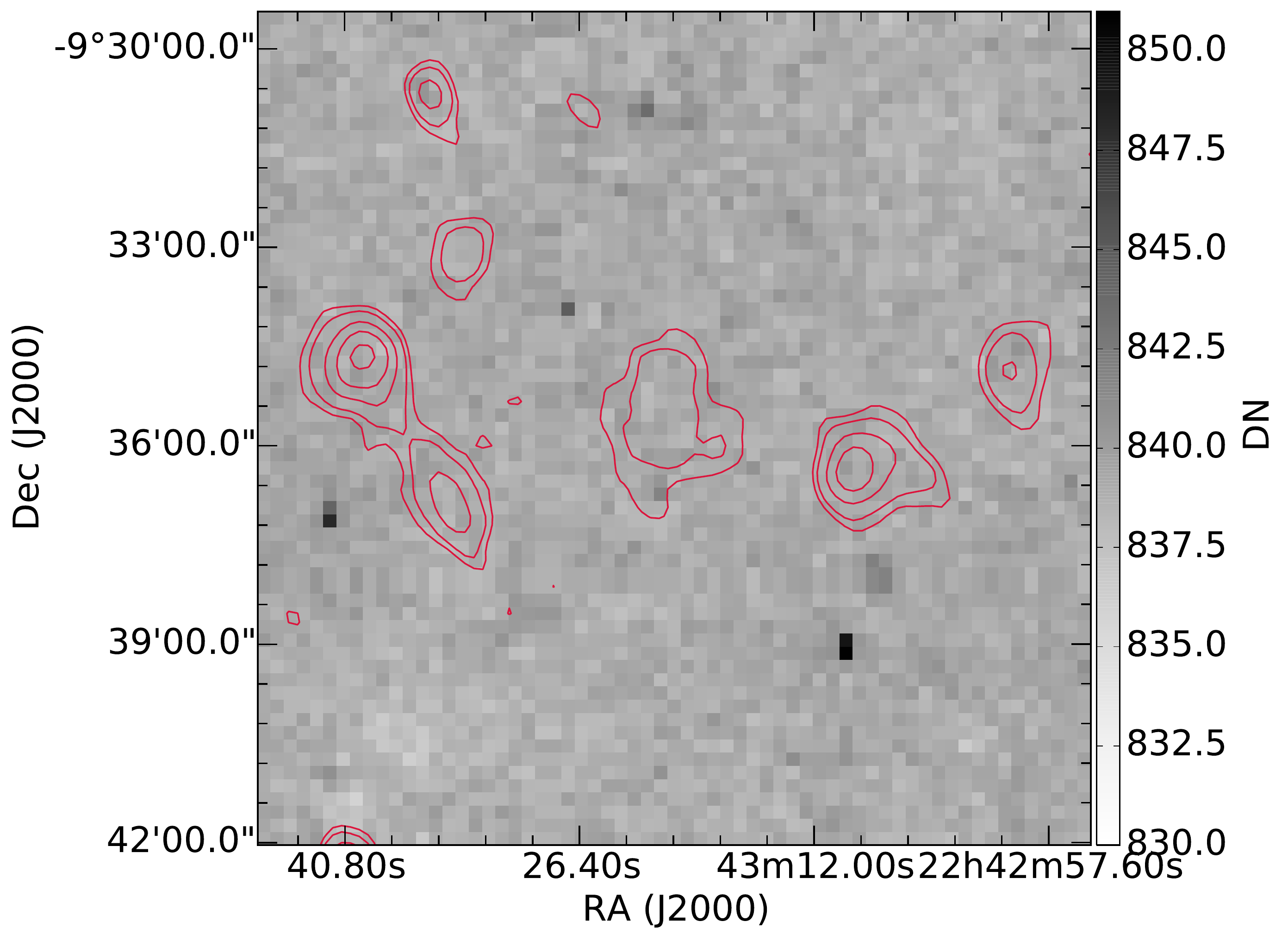}} & \subfloat[][]{\label{fig:X-ray_stars}}{\includegraphics[width=0.5\textwidth]{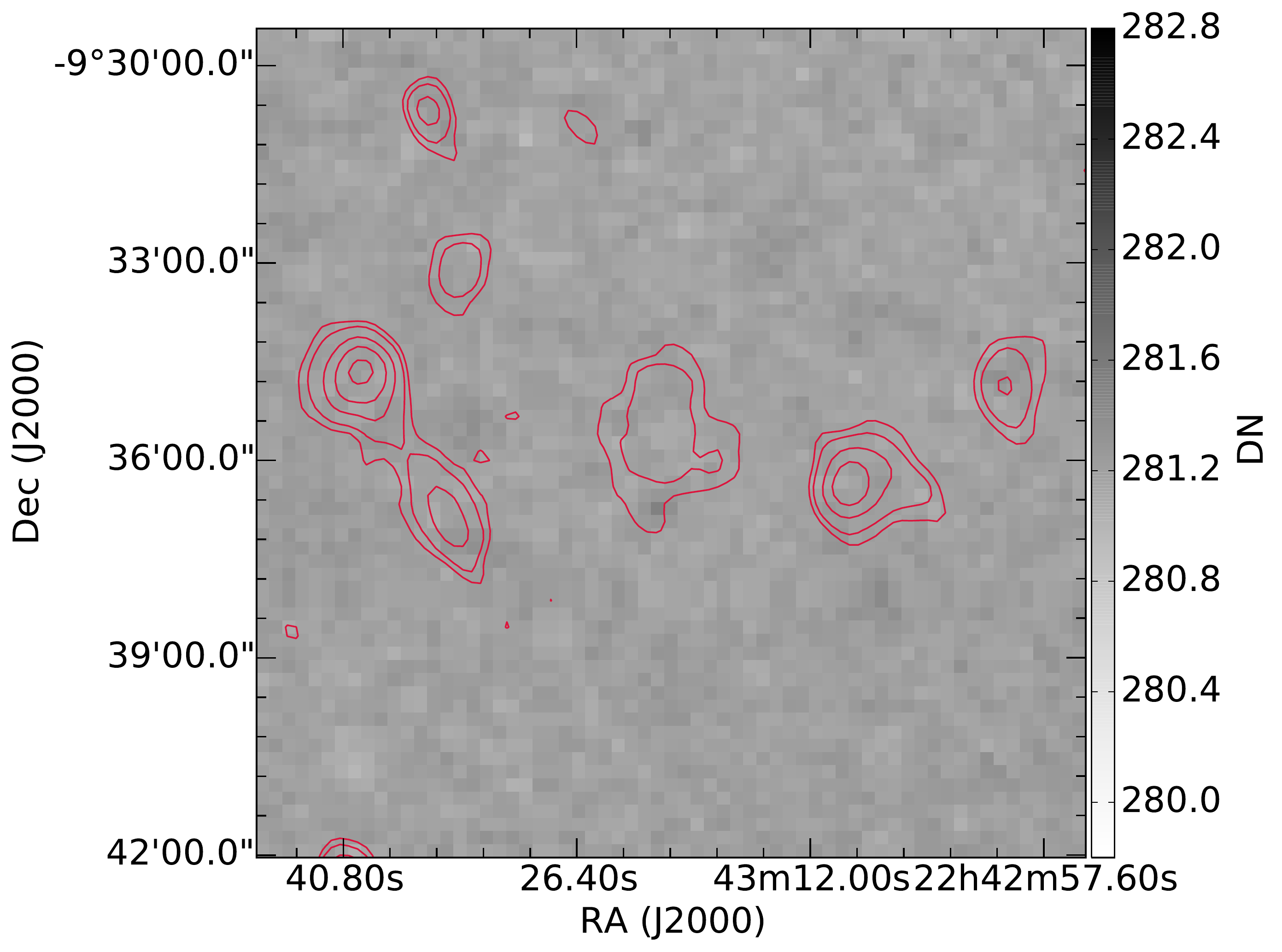}}\\
 \end{tabular}
     \caption{These images show MACS J2243.3-0935 at different wavelengths with red contours overlaid showing the tapered image. Contours are at 3, 5, 10, 15, 20 $\times$ $\sigma_{\rm rms}$ where $\sigma_{\rm rms}=200\, \rm \mu Jy/beam$. (a) IRIS 25 $\mu$m (b) IRIS 60 $\mu$m (c) IRIS 100 $\mu$m (d) SHASS H$\alpha$ (e) WISE 12 $\mu$m (f) WISE 22 $\mu$m}
     \label{figure:RegionD_multwavelength}
\end{figure*}

In order to rule out Galactic foreground emission as an explanation for region D, Figure~\ref{figure:RegionD_multwavelength} shows region D at multiple wavelengths. There is no significant emission in IRIS, SHASSA H$\alpha$, WISE or \textit{Planck}, and we conclude that region D is unlikely to be associated with Galactic foreground emission. 

Given the coincidence of region D with a filament, the radio emission might alternatively be associated with the WHIM rather than a radio relic. There is no agreement on precise predictions for magnetic fields in filaments but estimates range from 10$^{-4}$ to 0.1 $\mu$G \citep{Dolag1999,Bruggen2005,Ryu2008}. In section~\ref{sec:equiB} the equipartition magnetic field estimates for region D for a spectral index of 0.7 are greater than the estimates for magnetic fields in the WHIM. \citet{Araya-Melo2012} model cosmic rays in large scale structure and predict a flux density of 0.12 $\mu$Jy/beam at a frequency of 150 MHz for a 10 arcsec$^{2}$ beam. This is much lower than the flux density measured in region D or in other possible radio detections of the WHIM \citep{Bagchi2002,Farnsworth2013}. The high flux density and magnetic field estimates for region D suggest that it is unlikely to be a radio detection of the WHIM. 

\section{Conclusion}

We have discovered a radio halo in the merging cluster MACS J2243.3-0935 using GMRT observations at 610 MHz and KAT-7 observations at 1822 MHz. The radio halo has an integrated flux density of $10.0\pm2.0\, \rm mJy$, an estimated radio power at 1.4 GHz of $\left( 3.2 \pm 0.6\right) \times 10^{24}\, \rm W\,Hz^{-1}$ and a LLS of approximately $0.92\, \rm Mpc$. We calculated the equipartition magnetic field in the region of the halo for a range of $\alpha$ and $K_{0}$ values and find that the equipartition magnetic field is of order $1\, \rm \mu G$. Assuming a spectral index of $\alpha = 1.4$, the halo in MACS J2243.3-0935 lies on the empirical scaling relations observed for radio halos. 

We also detected a potential radio relic candidate to the west of the cluster. The candidate relic has a integrated flux density of $5.2\pm 0.8\, \rm mJy$, an estimated radio power at 1.4 GHz of $\left( 1.6 \pm 0.3\right) \times 10^{24}\, \rm W\,Hz^{-1}$ and a LLS of $0.68\, \rm Mpc$. The presence of a radio relic in MACS J2243.3-0935 would make this one of only a handful of clusters that host both a halo and a relic. Due to the position of the relic candidate on the outskirts of the cluster, where a filament meets the cluster, we conclude that the candidate is consistent with an infall relic. We rule out the possibility of the emission being associated with the WHIM in a filament as the measured flux density and estimated equipartition magnetic field strength are both much larger than expected values for the WHIM. We also exclude foreground galactic emission as an explanation as there is no significant emission in IRIS, SHASSA H$\alpha$, WISE or \textit{Planck}.

\label{sec:conclusion}

\section*{Acknowledgments}

 A.M.Scaife  gratefully  acknowledges  support  from  the  European  Research  Council  under grant ERC-2012-StG-307215 LODESTONE. The research of J.L.Han and Z.L. Wen are supported by the National Natural
Science Foundation of China (No. 11473034 and 11273029) and by the Strategic
Priority Research Program ``The Emergence of Cosmological Structures''
of the Chinese Academy of Sciences, Grant No. XDB09010200. We thank the staff of the GMRT who have made these observations possible. GMRT is run by the National Centre for Radio Astrophysics of the Tata Institute of Fundamental Research. We thank the staff of the Karoo Observatory for their invaluable assistance in the commissioning and operation of the KAT-7 telescope. The KAT-7 is supported by SKA South Africa and the National  Science  Foundation  of  South  Africa. We thank the anonymous referee for their useful comments that improved the content of this paper.




\bibliographystyle{mnras}
\bibliography{MACSJ2243} 








\bsp	
\label{lastpage}
\end{document}